\documentclass[twocolumn,secnumarabic, amsmath, amssymb, nobibnotes, nofootinbib, aps, prb, superscriptaddress]{revtex4-2}
\usepackage{graphicx}

\newcommand{\NEELAddress}{Univ. Grenoble-Alpes, CNRS, Inst. NEEL, Grenoble, France}
\newcommand{\IRIGAddress}{Univ. Grenoble-Alpes, CEA, IRIG-MEM-L$\_$Sim, Grenoble, France}

\begin{document}
\title{Light-hole states in a strained quantum dot: \\
numerical calculation and phenomenological models}

\author{K.~Moratis}
\affiliation{\NEELAddress}
\affiliation{\IRIGAddress}
\author{J.~Cibert}
\email[]{joel.cibert@neel.cnrs.fr}
\affiliation{\NEELAddress}
\author{D.~Ferrand}
\affiliation{\NEELAddress}
\author{Y.-M.~Niquet}
\affiliation{\IRIGAddress}

\date{\today}

\begin{abstract}
Starting from the numerical solution of the 6-band \textbf{k.p} description of a lattice-mismatched ellipsoidal quantum dot situated inside a nanowire, including a spin Zeeman effect with values appropriate to a dilute magnetic semiconductor, we propose and test phenomenological models of the effect of the built-in strain on the heavy hole, light hole and exciton states. We test the validity and the limits of a description restricted to a ($\Gamma_8$) quadruplet of ground states and we demonstrate the role of the interactions of the light-hole state with light-hole excited states. We show that the built-in axial strain not only defines the character, heavy-hole or light-hole, of the ground state, but also mixes significantly the light-hole state with the split-off band's states: Even for a spin-orbit energy as large as 1~eV, that mixing induces first-order modifications of properties such as the spin value and anisotropy, the oscillator strength, and the electron-hole exchange, for which we extend the description to the light-hole exciton. CdTe/ZnTe quantum dots are mainly used as a test case but the concepts we discuss apply to many heterostructures, from mismatched II-VI and III-V quantum dots and nanowires, to III-V nanostructures submitted to an applied stress and to silicon nanodevices with even smaller residual strains.
\end{abstract}

\pacs{Valid PACS appear here}

\maketitle

\section{\label{Intro}Introduction}

Quantum dots (QDs) are studied intensively as single photon or entangled photon emitters for quantum information processing. They can also host a single carrier which can be used as a qubit, with the possibility of optical manipulation using a charged exciton as intermediate state \cite{DeGreve2013}. Another promising direction deals with spin qubits in semiconductors, particularly silicon, with the prospect of their electrical manipulation \cite{Loss1998, Maurand2016, Crippa2018}.

In the case of quantum dots for optics, most of the studies are performed on flat QDs obtained by the Stranski-Krastanov process or by droplet epitaxy: The ground state of holes in such quantum dots is of the heavy-hole type, as a result of the quantum confinement and also possible mismatch strain. This situation results from the general properties of the valence band of a semiconductor with the zinc-blende or diamond structure, near the center of the Brillouin zone \cite{Luttinger1955}: The spin-orbit coupling splits the p-like Bloch functions into a doublet (with symmetry $\Gamma_7$), which gives rise to the so-called split-off band (SO), and a quadruplet (symmetry $\Gamma_8$), which gives rise to the topmost heavy-hole (HH) and light-hole (LH) bands. The SO band is often considered as far enough away to be disregarded when dealing with the top of the valence band. The HH and LH bands split further under a perturbation with axial symmetry, such as an axial strain \cite{BirPikus}. The confinement in a quantum well can also be described \cite{Bastard1988} in terms of HH and LH bands.

A heavy hole shows very anisotropic properties \cite{BirPikus, Bastard1988, Kesteren1990, Fishman2010} with two characteristic features: a spin state such that only the component of the spin operator along the quantization axis exhibits non-vanishing matrix elements within the heavy-hole doublet; an orbital state such that the matrix elements of the dipole operator between the heavy hole and a conduction band electron all vanish for the dipole component along the quantization axis. In other words, the spin of a heavy hole is along the axis, and the dipole it forms with an electron in the conduction band lies within the perpendicular plane.

A light hole features no such systematically vanishing matrix elements, and its spin and dipole anisotropy is opposite to that of a heavy hole. The in-plane spin of the light hole is twice larger than the axial spin. The dipole it forms with an electron in the conduction band is such that the oscillator strength of an optical transition is four times larger for axial than for in-plane polarization. Hence, either the spin properties or the optical selection rules can serve to identify the HH or LH character of holes through their anisotropy. From a more practical point of view, due to the presence of these non-vanishing matrix elements in all directions, having a light hole as the ground state can be advantageous for the optical manipulation of the spin of a confined carrier \cite{Quinteiro2014, Kumar2016} or of a magnetic impurity inserted in the dot \cite{Reiter2011, Moldoveanu2016}.

Although most quantum dots fabricated up to now exhibit a HH ground state, the LH ground state was obtained experimentally in a gallium arsenide QD with a tensile stress applied in-plane by a piezoelectric device \cite{Huo2014, Zhang2015}: Then the applied strain must overcompensate the built-in strain and the confinement. The ground state is also LH in an elongated QD with a compressive lattice mismatch, as shown theoretically \cite{Niquet2008, Zielinski2013, Greif2018} and demonstrated experimentally in QDs embedded in a nanowire \cite{Jeannin2017}: Then both confinement and built-in strain stabilize a light-hole ground state. Other systems are the zinc-blende nanocrystals, where confinement \cite{Sinito2014} but also mismatch strain play a role if a shell is added to the crystal core. More complex, multi-step nanostructures have also been explored to fabricate elongated structures with a compressive mismatch, essentially to control the polarization of the emitted light \cite{Greif2018, Li2009}.

The insertion of the quantum dot in a nanowire offers other attractive features. Indeed, the fabrication of a QD by the Stranski-Krastanov process or by droplet epitaxy is usually followed by the etching of a waveguide to collect the emitted light more efficiently \cite{Claudon2010, Dousse2010}. Recent studies have demonstrated the interest of a more accurate positioning of the quantum dot in a waveguiding nanowire \cite{Reimer2012, Dalacu2012} and even the possibility to insert the structure into a complete optical circuit \cite{Haffouz2018, Zadeh2016, Mnaymneh2019}. Such nanowires containing a QD have been grown mostly from InP \cite{Reimer2012, Dalacu2012, Harmand2013, Jaffal2019}, but nanowire-QDs structures have also been demonstrated with other materials such as CdSe \cite{Bounouar2012} or CdTe \cite{Szymura2015, Jeannin2017}. Particularly sharp optical lines are obtained in the case of III-V materials.

In parallel, spin qubits in silicon have gained much interest. This includes hole qubits \cite{Maurand2016, Crippa2018, Venitucci2019}, which offer strong opportunities for a fast, all-electrical manipulation of the qubit by electric dipole spin resonance. Recent studies have shown that small ($\sim 0.2 \%$) strains may be enough to switch from the normal heavy-hole ground state to a predominantly light-hole ground-state with totally different magnetic anisotropy \cite{Venitucci2018}. Understanding the complex interactions between the neighboring HH, LH and SO bands in such systems is, therefore, of fundamental importance in order to assess the potential of hole qubits and to optimize their design.

Theoretical studies on the subject of quantum dots in nanowires are mostly numerical \cite{Niquet2008, Zielinski2013, Greif2018}. The LH character of a ground state is generally established by its orbital properties (with consequences for the light emission pattern and the optical manipulation). This viewpoint should be complemented by the spin properties, and the fine structure resulting from the electron-hole exchange. In all cases, it is important to take into account the proper symmetry of the system: In particular, $\langle001\rangle$ and $\langle111\rangle$-oriented systems are different \cite{Bester2005, Singh2009, Schliwa2009}. However generic properties exist and must be identified, and possibly modeled phenomenologically.

Our goal is twofold:
\begin{itemize}
  \item (1) to assess through a numerical study the properties of a hole confined in a strained quantum dot as a function of its aspect ratio. Our focus is on elongated ellipsoidal QDs with a compressive strain induced by a lattice-mismatched shell. We use the parameters of a cadmium telluride quantum dot in a zinc telluride nanowire in order to make the connection with ongoing experimental efforts \cite{Jeannin2017, Artioli2019, Ferrand2019}. In addition to investigating the switching from the HH to a LH ground state as the aspect ratio increases from below to above unity \cite{Niquet2008, Zielinski2013}, we calculate the oscillator strengths and the spin properties. In particular, giant Zeeman shifts in dilute magnetic semiconductors such as the cadmium manganese telluride alloys (Cd,Mn)Te \cite{Furdyna1988} are of the same order of magnitude as the splitting between a QD's confined levels, thus providing a realistic way to adjust the interactions between these levels.
  \item  (2) to identify generic mechanisms and test phenomenological Hamiltonians for simple models: The results for the tellurides can be extended to other semiconductors. In group IV, III-V and II-VI materials, the approach widely used to describe the ground state considers the quadruplet formed by the ground state and the first level of opposite type (for instance, HH ground state with the first LH excited state) as a more or less independent system \cite{Jeannin2017, Tonin2012, Plachka2018}: We discuss the conditions for which this approach can be justified. We show that, if the valence band offset between the dot and its shell is not large enough with respect to the valence band edge splitting induced by the built-in strain, strong interactions exist between the hole states of the same type, as suggested by the idea of "supercoupling" \cite{Luo2015}. And we show that even for a material with a large spin-orbit coupling, the mixing with the split-off states (only $2\%$ in weight) strongly modifies the spin properties, the oscillator strengths and the electron-hole exchange interaction in the LH exciton.
      \end{itemize}

Our paper is organized as follows. Section \ref{Method} gives details mostly about the numerical calculations. The calculated strain distributions are described in Section \ref{Strain} (with a comparison to the analytical model of Eshelby assuming isotropic materials). The hole states are described in the absence of Zeeman effect in Section \ref{NoField}, and with the effects of a spin Zeeman effect in Section \ref{Field}. Section \ref{LHSO} is devoted to the LH-SO mixing and its effect on spin properties, oscillator strengths and electron-hole exchange. Section \ref{Conclusion} discusses the main conclusions and provides some comparisons to experimental data. The appendices give details about the different Hamiltonians introduced in this paper (Appendix~\ref{app:Hamiltonian}), and details of fits (Appendix~\ref{app:fits}). Appendix~\ref{app:ehexchange} presents an extension of the electron-hole exchange models beyond the HH exciton, and Appendix~\ref{app:MatPara} lists the material's parameters.

\section{\label{Method}Method}
The main differences with previous studies are (1) the presence of a lattice-mismatched shell around the QD (while a purely axial heterostructure was considered in Ref.~\cite{Niquet2008}), (2) inclusion of the piezoelectric field due to the $\langle111\rangle$ orientation (while in Ref.~\cite{Zielinski2013} results were reported to be qualitatively similar for a QD in $\langle001\rangle$ and $\langle111\rangle$ oriented nanowires), and (3) the evaluation of spin properties.

The structural and electronic properties of the QDs in nanowires were
calculated numerically with the \textbf{k.p} module of the TB$\_$Sim code \cite{code}. The QD is an ellipsoid of length $L$ along the $z$ axis of the nanowire, and diameter $D$~=~8~nm in the perpendicular, $xy$ plane. It is located at the center of a cylinder-shaped shell of diameter 120~nm and height 40~nm, with periodic boundary conditions along $z$. The $z$ axis is the $[111]$ direction of the zinc-blende structure, $x$ the $[1\overline{1}0]$ direction and $y$ the $[11\overline{2}]$ direction. We have checked that using a twice longer computing cell (80~nm) does not change significantly the results even for the longest QDs considered in this work ($L$=20~nm).

The strains are first computed with a finite element discretization of
continuum elasticity equations. The effect of strains on the valence band states is described using the deformation potentials of the Bir and Pikus Hamiltonian \cite{BirPikus}, and the Poisson equation is solved for the resulting piezoelectric potential. Finally, the hole states are calculated with a six-band \textbf{k.p} model discretized on the same mesh, using the Burt-Foreman operator ordering \cite{Burt1992, Foreman1997}.

The numerical calculations are performed with the parameters of CdTe for the QD and ZnTe for the shell, see Appendix \ref{app:MatPara}.

We use the bulk values of the piezoelectric constants, $e_{14} = $0.03 C~m$^{-2}$ for both CdTe and ZnTe. For the sake of simplicity and to avoid being too specific, we ignore the non-linear character of the piezoelectric effect \cite{Andre1996, Bester2006b, Bester2006, BeyaWakata2011, Germanis2016}, as well as the screening by mobile charges (if any) \cite{Boxberg2010}. We also ignore possible deviations from the linear Bir-Pikus Hamiltonian \cite{Kadantsev2010a, Gawarecki2019}.

We consider two values for the valence band offset (VBO) between
the unstrained materials of the QD and of the shell: a small valence band offset, 20~meV, relevant for the CdTe-ZnTe system (the "shallow QD"); and for the sake of comparison with more common configurations, a large value, 200~meV (the "deep QD").

We add to our Hamiltonian a spin Zeeman effect which acts only on the spin $\textbf{S}$ of the valence electron, with no orbital contribution: $\mathcal{H}_{sZ}=2 E_{sZ}\textbf{b}.\textbf{S}$, where $\textbf{b}$ is a unit vector and $E_{sZ}$ the external parameter describing the intensity of the effect.

This can be realized experimentally by doping the semiconductor with magnetic impurities such as manganese, thus forming a dilute magnetic semiconductor where the so-called giant Zeeman effect takes place \cite{Furdyna1988}. Its mechanism is well documented: In a bulk dilute magnetic material such as Cd$_{1-x}$Mn$_x$Te, each Mn impurity carries a spin $5/2$. When the paramagnetic system formed by these magnetic impurities is submitted to a magnetic field along the direction $\textbf{b}$, it acquires a magnetization $\textbf{M}=M\textbf{b}$. The intensity $M$ can be tuned, from zero up to a saturation value $M^{sat}$, by tuning the applied field. The exchange interaction between the carriers and the ensemble of spin-oriented magnetic impurities then induces a giant Zeeman shift of the valence band, $E_{sZ}=E_{sZ}^{sat} M/M^{sat}$. The shift at saturation $E_{sZ}^{sat}$ depends on the Mn-hole exchange energy and the Mn concentration $x$: In II-VI semiconductors such as (Cd,Mn)Te \cite{Furdyna1988, Gaj1979, Gaj1994}, it can reach more than 40 meV. This is equivalent to the effect of a field of intensity 700~T on a spin $1/2$ with $g=2$, but it is observed for an applied magnetic field of only a few teslas, hence with negligible influence on the orbital degrees of freedom. Note that the same exchange Hamiltonian, with the same value of $E_{sZ}$, applies for both the $\Gamma_8$ (HH and LH) and $\Gamma_7$ (SO) states: This was checked experimentally \cite{Twardowski1980} even in (Cd,Mn)Te in spite of the large spin-orbit splitting, 0.9~eV, as the interaction with the magnetic impurities takes place with a $d$-level of Mn at 3.5~eV \cite{Bhattacharjee1983}, sufficiently remote from the $\Gamma_8$ and $\Gamma_7$ band edges.

In the following, in order to reveal the spin properties of the confined hole states, we will plot the calculated quantities as functions of $E_{sZ}$.

\section{Built-in strain} \label{Strain}

\begin{figure} [t]
\centering
\includegraphics [width=\columnwidth]{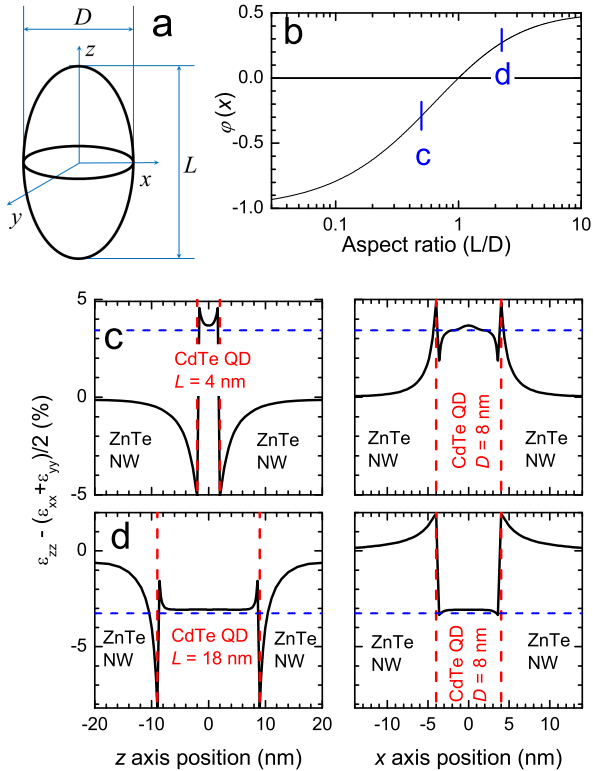}
\caption{(a) Scheme of the ellipsoidal CdTe inclusion, showing the axes, length $L$ and diameter $D$. Graph (b) shows the function $\varphi(\frac{L}{D})$ which represents the variation of the axial strain $\varepsilon_{zz}-\frac{1}{2} (\varepsilon_{xx}+\varepsilon_{yy})$ as a function of the aspect ratio $L/D$ in the Eshelby calculation, see Eq.~\ref{eq1} and \ref{eq2}. Graphs (c) and (d) are plots of the axial strain along the longitudinal axis $z$ and radial axis $x$. The solid lines show the result of the numerical calculation. The dashed blue lines are the Eshelby calculation. They are calculated for a flat QD, $L$=4~nm, in (c), and an elongated QD, $L$=18~nm, in (d), with diameter $D$=8~nm in both cases. The corresponding values of $L/D$ are indicated in graph (b).} \label{fig1}
\end{figure}

In this section, we discuss the strain distribution in the nanowires, namely the components of the strain tensor $\varepsilon_{\alpha\beta}$ with $\alpha,\beta=x,y,z$, as obtained from the numerical finite-element calculations. The elastic stiffness constants $c_{ij}$ and bulk lattice parameters $a_0$ of the materials are given in Appendix \ref{app:MatPara}. The lattice mismatch, $f=a_0^{shell}/a_0^{QD}-1$, is negative for a compressive mismatch, \emph{i.e.}, if the lattice parameter of the inclusion $a_0^{QD}$ is larger than that of the matrix $a_0^{shell}$. This is the case for a CdTe QD in a ZnTe matrix. The finite-element results will be used in the numerical calculation of the hole states discussed in the next sections.

As noted in Ref.~\cite{Rybchenko2007} for flat QDs, it is interesting to compare these results to the analytical formula obtained by Eshelby \cite{Eshelby1957} for the built-in strain in a perfectly ellipsoidal inclusion in an infinite matrix (Fig.~\ref{fig1}a), both having the same, isotropic stiffness tensor. The strain configuration in the inclusion is extremely simple: Only two components are non-zero, the isotropic (hydrostatic) strain $\frac{1}{3} (\varepsilon_{xx}+\varepsilon_{yy}+\varepsilon_{zz})$, and the axial strain $\varepsilon_{zz}-\frac{1}{2}(\varepsilon_{xx}+\varepsilon_{yy})$. Both strain components are uniform over the inclusion, and proportional to the lattice mismatch $f$. The axial strain depends on the aspect ratio $L/D$ of the inclusion, with a formula which can be extracted from Ref.~\cite{Eshelby1957} (and which is actually of the same form as that of the demagnetizing field or the polarizability of an ellipsoid):

\begin{equation}\label{eq1}
\varepsilon_{zz}-\frac{1}{2} (\varepsilon_{xx}+\varepsilon_{yy})=-\frac{1+\nu}{1-\nu}~\varphi(\frac{L}{D})~f
\end{equation}
where $\nu$ is the Poisson coefficient, and
\begin{eqnarray}\label{eq2}
\varphi(x)&=\frac{1}{2}-\frac{3}{2}\left[1-\frac{x \cos^{-1} (x)}{\sqrt{1-x^2}}\right] \frac{1}{1-x^2}\nonumber\\
\varphi(x)&=\frac{1}{2}-\frac{3}{2}\left[1-\frac{x \cosh^{-1} (x)}{\sqrt{x^2-1}}\right] \frac{1}{x^2-1}
\end{eqnarray}
for $x<1$ and $x>1$, respectively. $\varphi(x)$ varies from -1 for $x=0$ (the flat ellipsoid mimics a quantum well) to $\frac{1}{2}$ for $x\mapsto\infty$ (the core-shell nanowire), through 0 by symmetry for $x=1$ (when the ellipsoid is a sphere), see Fig.~\ref{fig1}b.

The two limits (quantum well and core-shell nanowire) also have analytical solutions if the symmetry axis is the $\langle111\rangle$ axis of a cubic material: $\varepsilon_{zz}-\frac{1}{2} (\varepsilon_{xx}+\varepsilon_{yy})=-3 \frac{c_{11}+2c_{12}}{c_{11}+2c_{12}+4c_{44}}f$ for the well-known case of the $(111)$ quantum well, and $\frac{c_{11}+2c_{12}}{c_{11}+c_{12}+2c_{44}}f$ for the $\langle111\rangle$ core-shell nanowire \cite{Ferrand2014}. Using the stiffness tensor of CdTe, the values are $3 \frac{c_{11}+2c_{12}}{c_{11}+2c_{12}+4c_{44}}=0.98$ and $\frac{c_{11}+2c_{12}}{c_{11}+c_{12}+2c_{44}}=0.51$, so that interpolating with $3 \frac{c_{11}+2c_{12}}{c_{11}+2c_{12}+4c_{44}} \varphi(x)$ for $x<1$ and $2\frac{c_{11}+2c_{12}}{c_{11}+c_{12}+2c_{44}} \varphi(x)$ for $x>1$ should provide a good estimate of the axial strain in the CdTe QD. For other semiconductors such as GaAs or germanium, the ratio between the two asymptotic limits remains close to the isotropic limit, $-2$, to within a few percent.

We will use the Eshelby formula to calculate analytically the trends of various quantities as a function of the aspect ratio, as shown in Figs.~\ref{fig2} and \ref{fig4}.

Parts (c) and (d) of Fig.~\ref{fig1} display the result of the finite-element calculation of the axial strain along the longitudinal axis $z$ and the radial axis $x$, in a flat QD ($D=$~8~nm and $L=$~4~nm) and in an elongated QD ($D=$~8~nm and $L=$~18~nm). The main features are the large value of the axial strain in the QD, quite close to the Eshelby value (dashed line), and the large mismatch-induced jump at the interface. The strain in the shell is strong in an area limited to the vicinity of the interface (as in a spherical inclusion \cite{Ithurria2007} or in a circular core-shell nanowire \cite{Ferrand2014}). Fig.~\ref{fig1} shows additional modulations: Small but visible in the dot, they exhibit a threefold symmetry around the $z$ axis (as shown in Ref.~\cite{Gronqvist2009}) and are due to the cubic symmetry of the stiffness tensor resulting from the zinc-blende crystal structure.

Note that the cubic symmetry is also present in the piezoelectric tensor, in the Bir-Pikus Hamiltonian, and in the Luttinger Hamiltonian.

The axial strain along the $\langle111\rangle$ axis induces a splitting $2Q=-\frac{2d}{\sqrt{3}}[\varepsilon_{zz}-\frac{1}{2} (\varepsilon_{xx}+\varepsilon_{yy})]$ between the light hole and the heavy hole at the valence band-edge. The strain present in an elongated quantum dot with a compressive lattice mismatch is such that $2Q<0$, so that the ground state of the dot is formed on the light hole band. For the dot of Fig.~\ref{fig1}d, we calculate $2Q=-190$~meV with the Eshelby approach, and -180~meV at the center of the dot according to the numerical calculation. For a flat dot, $2Q>0$ and the ground state of the dot is formed on the heavy hole band. For the dot of Fig.~\ref{fig1}c, we calculate $2Q=200$~meV with the Eshelby approach, and 220~meV at the center of the dot according to the numerical calculation. Figure~\ref{fig1}b shows that the splitting $2Q$ decreases monotonously when the ratio $L/D$ increases.

\section{\label{NoField} Hole states}

Figures \ref{fig2} and \ref{fig3} present the energy and envelope functions calculated for the first six Kramers doublets in a "deep" and a "shallow" quantum dot. The QD's diameter is $D=8$~nm, and the length $L$ varies from 2 to 20~nm. The unstrained valence band offset is set to 200 meV for the deep QD, 20 meV for the shallow QD; All other parameters are those of CdTe embedded in ZnTe.

\subsection{"Deep" Quantum Dot} \label{Deep}

 \begin{figure} [!]
\centering
\includegraphics [width=\columnwidth]{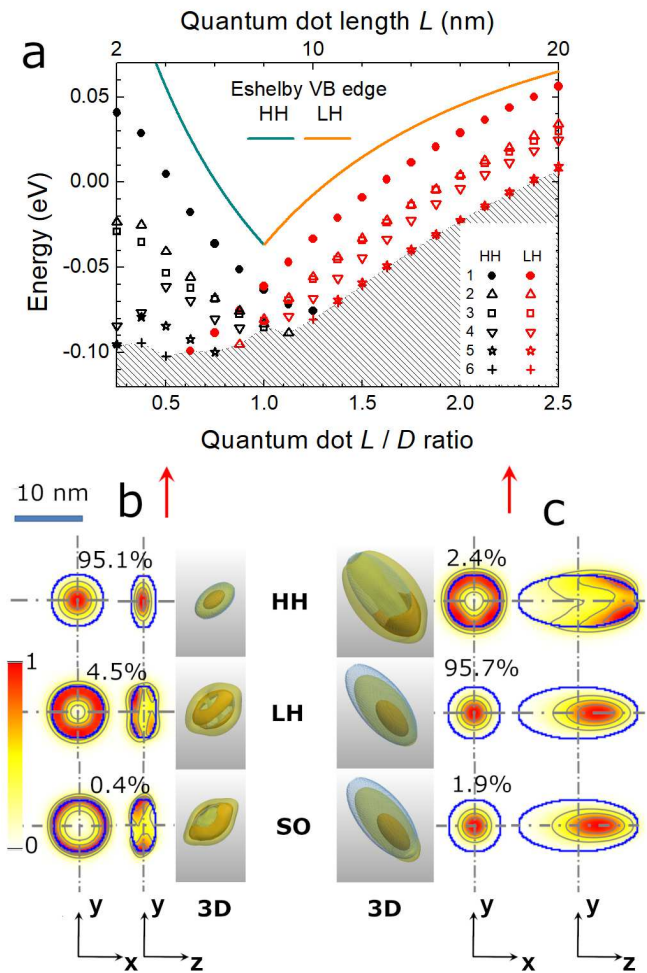}
\caption{Deep QD (valence band offset 200 meV at zero strain): Part (a) shows the energy of the first 6 levels as a function of the length/diameter ratio. The zero of energy is the top of the valence band of unstrained bulk CdTe. The valence band edge in the QD calculated using the Eshelby formula is indicated by colored lines (see text). Additional excited states, not calculated, are in the hatched zone. Closed symbols mark the lowest-lying HH and LH states. The color of symbols indicates the nature of the main hole component (red for LH, black for HH). Figure parts (b) and (c) show normalized projections onto the $xy$ and $yz$ planes (integrated over the normal direction $z$ and $x$ respectively). They also show 3D iso-surfaces of the envelope function probabilities for the HH, LH and SO components of the ground state of a QD. Part (b) is for a flat QD, $L$=4 nm, $D$=8 nm. Part (c) is for an elongated QD, $L$=18 nm, $D$=8 nm. Scale of the projections is as indicated. On the 3D plots, yellow is the 0.05 iso-surface, and orange is the 0.5 iso-surface, as measured with respect to the maximum.} \label{fig2}
\end{figure}

Figure \ref{fig2}a shows the energy diagram for the deep QD. The top of the valence band obtained from the Eshelby approach is given by the solid line. It evidences the crossing between the LH and HH bands at $L/D=1$. In a simple approach, this line indicates the bottom of the QD potentials which confine the light holes and heavy holes. As expected from confinement, the ground state follows the corresponding band edge, with a shift which increases when $L$ decreases. This remains nonetheless a qualitative trend: In the real QD, the valence band edge is not uniform as it includes the effects of the piezoelectric field and of non-uniform strain components.

Figure~\ref{fig2}b displays 3D plots of the ground state envelope functions for a flat QD at the value $L/D=0.5$ marked by an arrow in Fig.~\ref{fig2}a, and the corresponding projections onto the $xy$ and $yz$ planes. These are not cross-sections: For instance, the projections onto the $xy$ plane are obtained by integrating the square of each envelope function over the $z$ axis, and normalizing. Figure~\ref{fig2}b demonstrates a HH character with a weight of 95\%, and an s-like envelope function. There is also however a LH component: It features an envelope function of higher-order (note the nodes on the $z$ axis and in the $xy$ equatorial plane), as expected for a system with an approximately circular symmetry which couples states with the same projection of the total angular momentum (including that of the envelope function). There is also a small (0.4\%) SO component.

Figure~\ref{fig2}c displays the same quantities for an elongated QD, with a principal LH contribution with an s-like envelope function, a small HH contribution with a higher-order envelope, and a significant (2\%) SO contribution also with an s-like envelope function. The envelope functions are shifted along the axis as a result of a piezoelectric field principally oriented along $z$.

The HH and LH ground states cross at $L/D\simeq1$ and each of them can be followed easily on both sides of $L/D=1$. This is true also for the first excited states, represented by open symbols. The principal envelope function of each of these excited states has a node (not shown); For instance, in the elongated QD, the first three, almost degenerate levels feature p-type envelopes.

Figure~\ref{fig4} displays the weight of each component, LH, HH and SO, in the ground state, and the relative oscillator strengths of the transition to the electron ground state, as a function of the aspect ratio $L/D$. These quantities are calculated assuming a thermal distribution over the six Kramers doublets, at a temperature equal to 4K. Note that the oscillator strength is normalized, \emph{i. e.}, we plot $P_\alpha^2=\overline{|\langle p_\alpha\rangle|^2}~/~\sum_{\alpha=x,y,z}\overline{|\langle p_\alpha\rangle|^2}$, where $\langle p_\alpha \rangle$ is the valence to conduction band matrix element of the momentum operator along direction $\alpha$, see Eq.~\ref{dipole}, and $\overline{O}$ denotes a thermal average. Some LH-HH mixing is visible in Fig.~\ref{fig4}a, mainly around $L/D=1$. It involves envelope functions of higher order, which are orthogonal to the s-like envelope function of the electron's ground state: As a result, the mixing gives no sizable contribution to the normalized oscillator strengths, Fig.~\ref{fig4}b, which mimic those of a pure HH exciton ($P_x^2+P_y^2=1$, $P_z^2=0$) when $L<D$, or those of a pure LH exciton ($P_x^2+P_y^2=\frac{1}{3}$, $P_z^2=\frac{2}{3}$) when $L$ is just larger than $D$.

An unexpected result is the evolution of the oscillator strengths when increasing the aspect ratio further: They deviate significantly from those of a pure LH. This is due to a LH-SO mixing, noticeable in Fig.~\ref{fig4}a, which will be discussed later on (Section \ref{LHSO}).

\subsection{"Shallow" Quantum Dot} \label{Shallow}

\begin{figure} []
\centering
\includegraphics [width=\columnwidth]{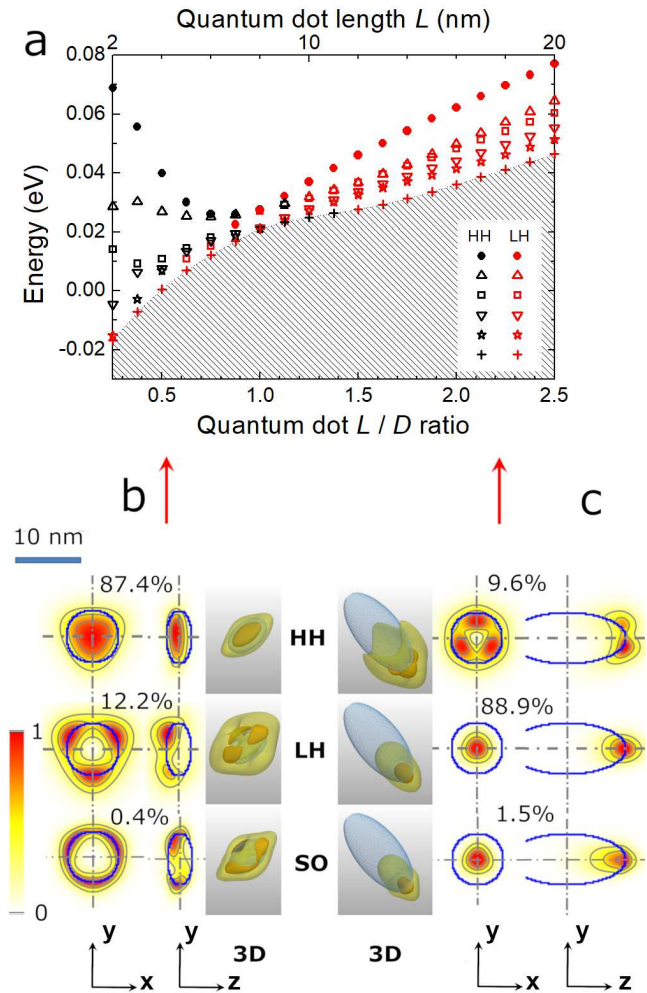}
\caption{Same as Fig. \ref{fig2} for a shallow QD (valence band offset 20 meV at zero strain).} \label{fig3}
\end{figure}

Figures \ref{fig3} shows the same results for the shallow QD. When compared to the deep QD case, as expected:
\begin{itemize}
  \item The energy range is smaller.
  \item  The envelope functions leak out from the QD. This leakage is particularly strong for the second component (LH for the flat dot and HH for the elongated dot) which is almost expelled from the QD into the shell, and acquires features related to the three-fold symmetry of the cubic system around $\langle111\rangle$. The overlap of this envelope with the envelope of the main hole component, and its overlap with the envelope of the electron, are vanishingly small, so that it plays a minor role in the properties discussed in this paper. Note also the shift along $z$ of the LH state of the elongated dot, which is induced by the axial piezoelectric field: It is much more visible in the present, shallow QD, where it induces a significant leak into the barrier, than in the deep QD, where the leak is limited by the large barrier height.      \end{itemize}

\begin{figure} []
\centering
\includegraphics [width=\columnwidth]{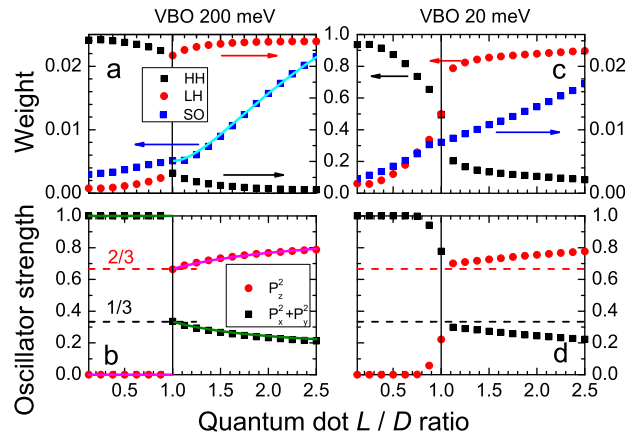}
\caption{Fig.~(a) Weights of the three components, LH (red symbols), HH (black symbols) and SO (blue symbols), and Fig.~(b) the relative longitudinal and transverse oscillator strengths for a deep QD (valence band offset VBO=200 meV at zero strain). Figs.~(c,d) show the same for a shallow QD (valence band offset VBO=20 meV at zero strain). Symbols display the results of the numerical calculation, solid curves in (a) and (b) are from the analytical calculation of the strain-induced LH-SO mixing, see Section \ref{LHSO}.} \label{fig4}
\end{figure}

But also:
      \begin{itemize}
  \item  Instead of the clear-cut level-crossing observed for the deep QD at $L/D=1$, the energy diagram suggests the presence of a sizable anticrossing at this position,
  \item  This is confirmed by the strong LH-HH mixing observed even far from $L/D=1$, see Fig.~\ref{fig4}c.
\end{itemize}

Similar trends will be observed when we add the spin Zeeman Hamiltonian in the next section.

\section{\label{Field} Effects of the Spin Zeeman Hamiltonian }

In this section we add the spin Zeeman Hamiltonian in the QD, and not in the surrounding material. This is realized experimentally by a (Cd,Mn)Te dot in a ZnTe nanowire \cite{Szymura2015, Jeannin2017}. In addition, the structure comprises an additional (Zn,Mg)Te shell, in order to isolate the QD from defects at the sidewalls of the nanowire. This additional shell presents a small tensile lattice mismatch with respect to the inner ZnTe shell, so that the axial strain in the QD is modified and the switching between HH and LH ground states is displaced from $L/D=1$, as demonstrated in Ref.~\cite{Artioli2019}. We checked that the effect of the additional strain far from $L/D=1$ is much smaller: It is not significant for the different aspects discussed in the present paper at the values of the valence band edge that we consider.

\begin{figure*} [t]
\centering
\includegraphics [width=1.6\columnwidth]{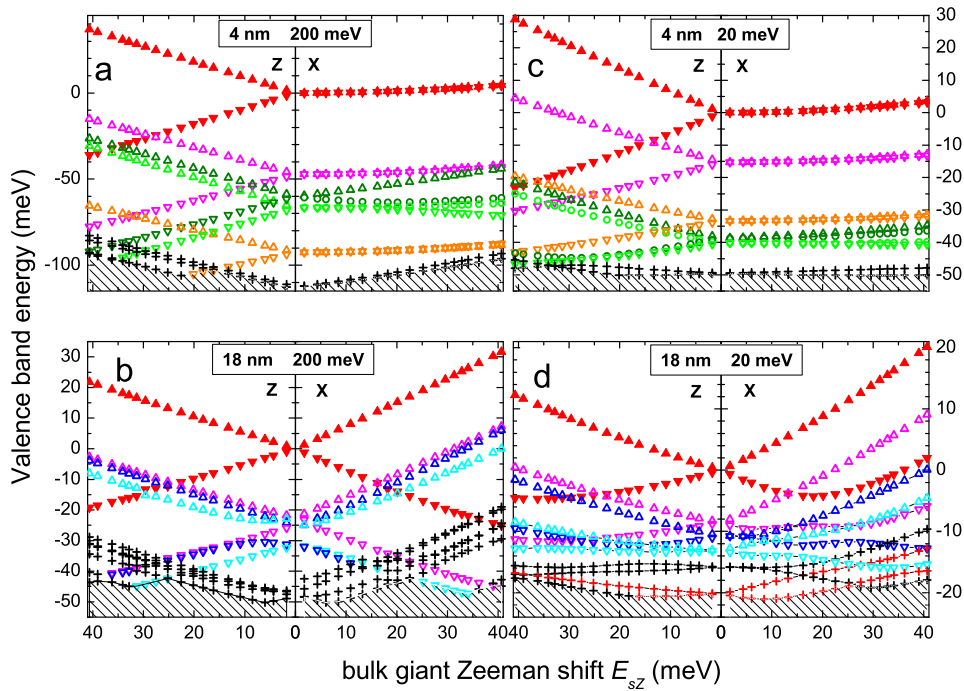}
\caption{Spin Zeeman effect calculated for various shapes of quantum dots. (a) the deep flat QD, (b) the deep elongated QD, (c) the shallow flat QD, and (d) the shallow elongated QD. Different symbols are guides for the eye. The horizontal scale is the giant Zeeman shift of a hole in bulk (Cd,Mn)Te.} \label{fig5}
\end{figure*}

Figure~\ref{fig5} shows the position of the first 12 levels for the four types of QDs described in Section \ref{NoField}: deep and shallow, flat and elongated. In the left panels, the magnetic field is longitudinal, \emph{i.e.}, applied along the nanowire axis ($z$ axis). In the right panels, it is transverse, \emph{i.e}., applied in the normal plane ($x$ axis). The horizontal scale is the spin Zeeman shift $E_{sZ}$ of the bulk dilute magnetic semiconductor.

We now discuss the main features of the spin Zeeman Hamiltonian within the $\Gamma_8$ states. We show (\Ref{LHvsHH}) that the spin properties confirm the identification of the HH and LH states, including the LH-HH coupling due to a transverse field. Then (\Ref{Confinement}) we describe the interplay between the spin Zeeman effect and confinement in the dot. We also highlight how the spin Zeeman effect reveals and controls the interactions between the LH ground state and LH excited states (\Ref{LHLH}). Finally (\ref{spinSummary}) we show that a quantitative agreement cannot be achieved when considering only the $\Gamma_8$ states: The following section (\Ref{LHSO}) shows how this is achieved by considering the effect of the split-off states.

\subsection{Light hole vs. heavy hole}\label{LHvsHH}

The main features confirm the expectations from Section \ref{NoField}.

In the "deep", elongated QD, we recognize the typical LH behavior, with a larger Zeeman splitting under a transverse magnetic field than under a longitudinal one (for a pure light hole state, we expect a spin $\frac{1}{3}$ in the plane and $\frac{1}{6}$ along the axis).

For the "deep", flat QD, we calculate a large Zeeman splitting under a longitudinal magnetic field, with a slope close to that of a spin $1/2$ (see a zoom in Fig.~\ref{fig6}a), and a (quasi)-absence of Zeeman effect for a transverse field. This is found for the ground level (closed symbols) and for the first excited levels (open symbols). These are the signature of HH states.

The transverse field has nonetheless a small but measurable effect on the HH states, as shown by the zoom in Fig.~\ref{fig6}b: Such a shift is usually interpreted \cite{Plachka2018, Jeannin2017, Artioli2019} as the result of the coupling induced by the transverse field between the HH state and LH states with the same type of envelope function (s-like) and a significant probability of presence within the dot. Indeed, the average shift is quadratic and the splitting is cubic in the bulk Zeeman shift $E_{sZ}$, as expected within a $\Gamma_8$ quadruplet with a large HH-LH splitting. Details of the fit in Fig.~\ref{fig6}b are given in Appendix~\ref{app:fits}: The position of the excited LH component agrees reasonably well with the expected LH-HH splitting ($2Q$). However, there is no evidence that a single LH doublet is involved.

\subsection{Confinement } \label{Confinement}

We discuss here two examples of the effect of an incomplete confinement of the hole in the QD.

Figure~\ref{fig6} is a zoom into Fig.~\ref{fig5}a, showing the ground Kramers doublet of the deep, flat QD. The Zeeman shift with the magnetic field along $z$ is smaller than $E_{sZ}$, hence smaller than expected for the spin $\frac{1}{2}$ of a heavy hole (dashed line). Actually the calculation shows that for this state (see Fig.~\ref{fig2}), the weight of the HH component is $95\%$, and furthermore that the envelope function of this HH component is only $95\%$ confined in the QD, with the remaining $5\%$ in the shell where the spin Zeeman shift is zero. The rest of the state is mostly (weight $4\%$) LH, with a smaller spin and a low probability of presence ($30\%$) within the QD where the spin Zeeman shift is acting, so that its contribution to the Zeeman effect is negligible. Taking these two reductions into account leads to an effective spin $\simeq 0.45$. The corresponding Zeeman shift shown by the solid lines in Fig.~\ref{fig6}a is in good agreement with the numerical calculation (symbols).

Another, more dramatic example is given in Fig.~\ref{fig7}: the case of the shallow elongated dot. The ground state is a LH state confined in the dot to about 60\% at zero field. But as the Zeeman shift takes significant values with respect to the valence band offset, the confinement is significantly altered. It increases to more than 70\% for the spin-up ground state, Fig.~\ref{fig7}a, and decreases for the spin-down excited state, not shown. Moreover, the change is larger when the Zeeman effect is larger, \emph{i.e}, larger for a transverse than for a longitudinal  magnetic field. This linear variation of the probability of the hole's presence in the dot where the spin Zeeman effect is active contributes to a quadratic component of the Zeeman shift, visible in Fig.~\ref{fig5}d and in the zoom (Fig.~\ref{fig7}b). Two other mechanisms which also contribute to this quadratic component are discussed in the next subsection, \ref{LHLH}, and in Section \ref{LHSO}.

\subsection{Mixing between light-hole states}\label{LHLH}

Level anticrossings are well developed and visible in the energy plot of the shallow elongated QD, see Fig.~\ref{fig5}d. For the ground doublet (see Fig.~\ref{fig7}b which is a zoom into Fig.~\ref{fig5}d), the anticrossing develops over a range narrow enough that it is fully scanned by the transverse magnetic field. The anticrossing is also partially visible with the axial magnetic field (see Fig.~\ref{fig5}d). The slopes on both sides of the anti-crossing are characteristic of LH states. The mechanism of the previous section (LH-HH mixing induced by the spin Zeeman effect) is therefore ruled out. This suggests instead a weak interaction with one (or several) nearby LH levels.

A good fit (curves in Fig.~\ref{fig7}b, details are given in appendix~\ref{app:fits}) is obtained by considering two interacting states: the LH aground state, and an excited LH state, which is reasonably well confined in the QD, about 15 meV from the ground state. A closer examination of Fig.~\ref{fig5}d shows that several levels are located in the relevant energy range: The ground state anti-crosses a group of levels near that energy rather than a single, well-defined state.

If we consider only the upper spin branch, the effect of the anticrossing is well represented by a quadratic dependence on $E_{sZ}$, which adds to the effect due to confinement described in Section \ref{Confinement}.

It is interesting to note that, compared to the elongated shallow dot, no such anticrossing is observed for the deep elongated dot (Fig.~\ref{fig5}b), in spite of a very similar distribution of excited levels. That suggests that the interaction between the ground doublet and the excited LH levels takes place in the shell. Lattice-mismatched QDs indeed display non-uniform strain configurations near the interface and in the shell, giving rise to specific terms in the Bir-Pikus Hamiltonian (usually labelled $R$ and $S$). The cubic symmetry also plays a role as it affects the strain configuration through the anisotropy of the compliance coefficients, and the Bir-Pikus and Luttinger Hamiltonians: Its effect is visible in the envelope function projections in Fig.~\ref{fig2}b and c and even more in Fig.~\ref{fig3}b and c. As such, the present LH-LH anticrossing is revealed by the applied magnetic field but it involves non-magnetic couplings between envelope functions of various symmetries: It appears quite different from the HH-LH anticrossing, which results from a coupling induced by the applied field itself, and takes place between states with a strong overlap of the envelope functions.

\subsection{Summary on HH-LH spin properties}\label{spinSummary}

To sum up, most of our calculated spin properties agree with the general expectations for a heavy hole or light hole, including the anticrossing induced by the transverse magnetic field between LH and HH confined in the QD \cite{ Plachka2018, Jeannin2017, Artioli2019, Luo2015}. Deviations are attributed to a probability of presence in the QD less than unity, and the occurrence of LH-LH anticrossings that we relate to the non-zero probability of presence of the hole in the shell, where the strain distribution is non-uniform.

However, the agreement is not quantitative. The ground state in the deep, elongated QD is essentially of LH character. For a pure LH, we expect a spin value $\frac{1}{6}$ along the $z$ axis, and $\frac{1}{3}$ along the $x$ axis. As confinement is not total (see Fig.~\ref{fig2}), we expect even slightly smaller values of the spin, with the same anisotropy. Figure~\ref{fig8}a shows that the shifts calculated numerically (symbols) are definitely larger than those of a LH (dashed lines), and have a smaller anisotropy. These higher values cannot be attributed to the presence of the small HH component visible in Fig.~\ref{fig2}, as its envelope function is quite delocalized out of the QD, and is thus mostly insensitive to the spin Zeeman effect present only in the QD. We show now that a quantitative agreement is obtained if we include the effect of the split-off states.

\begin{figure} []
\centering
\includegraphics [width=\columnwidth]{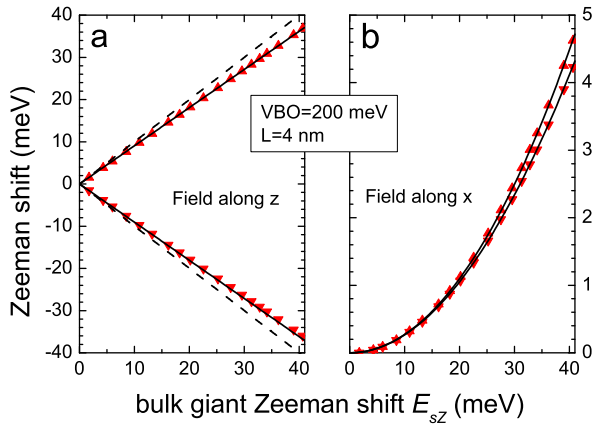}
\caption{Details for the flat, deep QD (a zoom into Fig.~\ref{fig5}a): Panel (a) shows the Zeeman shift of the ground Kramers doublet with magnetic field along $z$. The dashed line is for pure HH character with spin $\frac{1}{2}$. Symbols are from numerical calculations. Solid lines show the contribution of the part of the HH component located in the QD. Panel (b) is for the magnetic field along $x$. The solid curves are calculated as explained in appendix~\ref{app:fits}.} \label{fig6}
\end{figure}

\begin{figure} []
\centering
\includegraphics [width=\columnwidth]{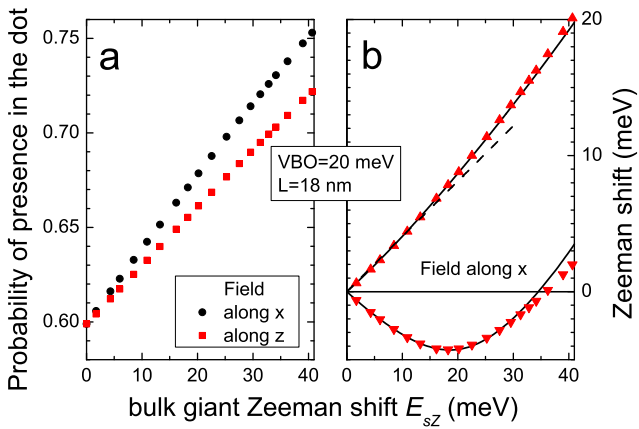}
\caption{Details for the shallow, elongated QD. Panel (a) shows the probability of presence of the ground state in the QD. Panel (b) shows the Zeeman shift of the ground Kramers doublet. Symbols are numerical data, solid curves are from the analytical calculation including LH-SO mixing, reconfinement and LH-LH anticrossing as described in the text. The dashed line shows the slope at low field.} \label{fig7}
\end{figure}

\section{Light-hole split-off mixing} \label{LHSO}

In this section we show that in spite of a small weight of the split-off (SO) component in the light hole states, the mixing induced by the axial strain significantly alters the oscillator strengths and the spin values. A simple analytical expression is tested. Finally, this analytical approach is applied to another system: a III-V compound, flat quantum dot, submitted to a biaxial strain \cite{Huo2014}.

\begin{figure} []
\centering
\includegraphics [width=\columnwidth]{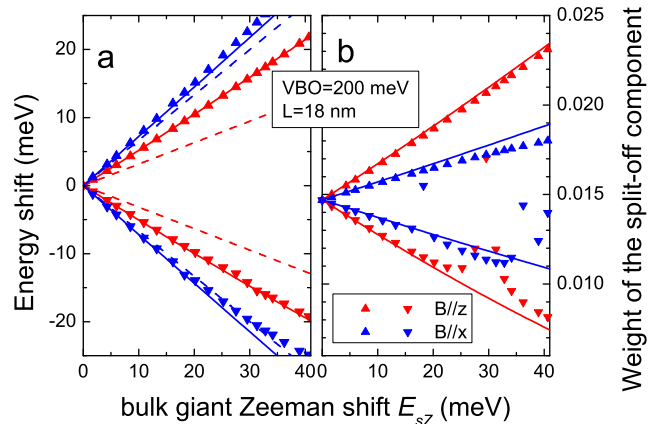}
\caption{Light-hole / split-off states mixing and its effect on the Zeeman shift. Panel (a), Zeeman shift of the ground Kramers doublet of the elongated, deep QD. Panel (b), weight of the SO component. Symbols are from numerical calculation. Dashed lines in (a): pure LH Zeeman shift (spin $\frac{1}{3}$ along $x$ and $\frac{1}{6}$ along $z$). Solid lines: analytical calculation of the LH-SO mixing using Eqs.~\ref{LHSOmatrixp} to \ref{100LHSOmatrixm}.} \label{fig8}
\end{figure}

Indeed, the calculated LH states contain a SO component: Its weight depends on the form factor, see Fig.~\ref{fig4}a, and on the spin Zeeman effect, Fig.~\ref{fig8}b. It is small, a few \% at most, but we will see that its effect is determined by the mixing amplitude, \emph{i.e.}, it is proportional to the square root of the weight shown in Fig.~\ref{fig4}a and Fig.~\ref{fig8}b. Moreover, both the LH and SO components of the mixed hole state have a mostly s-like envelope function (Fig.~\ref{fig2}c) and thus directly contribute to matrix elements such as those of the dipole with the electron state (the oscillator strengths in Fig.~\ref{fig4}c) or the spin Zeeman shift (Fig.~\ref{fig8}a).

The axial strain in the elongated dot is uniform to a good approximation. A simple hypothesis is that the LH ground state is mixed with a SO ground state with the same envelope function\footnote{Actually, the LH may also be mixed with an ensemble of SO states with energies $\sim\Delta_{SO}$, each with a different envelope function confined (or not) in the dot. The coupling between an isolated LH state and such a dense set of remote SO states tends to imprint the LH envelope onto the admixed SO states, so that the LH and SO envelopes end up very similar. Another example was discussed in Section \ref{LHLH}.}, well-confined in the SO potential, at an energy around $\Delta_{SO}$. Then we may calculate the effect of strain as in bulk material: The Bir-Pikus Hamiltonian $\mathcal{H}_Q$ due to the axial strain has non-vanishing matrix elements between the LH and the SO band edges, which are given in appendix \ref{app:Hamiltonian}, Eq.~\ref{HQ}.

It is helpful to write the mixing to first order in $\eta=Q/\Delta_{SO}$. The mixed LH states are $\widetilde{|\Gamma_8,\frac{1}{2}\rangle}=|\Gamma_8,\frac{1}{2}\rangle -\eta\sqrt{2}|\Gamma_7,\frac{1}{2}\rangle$
 and
 $\widetilde{|\Gamma_8,-\frac{1}{2}\rangle}=|\Gamma_8,-\frac{1}{2}\rangle +\eta\sqrt{2}|\Gamma_7,-\frac{1}{2}\rangle$.
When developed using the expressions for the LH and HH states, Eq.~\ref{States}, we have:

\begin{eqnarray}\label{MixedStates}
\widetilde{|\Gamma_8,\frac{1}{2}\rangle}&&=\frac{\sqrt{2}}{\sqrt{3}}(1-\eta)|iZ\rangle|+\rangle\nonumber\\
&&-\frac{1}{\sqrt{6}}(1+2\eta)[|iX\rangle+i|iY\rangle]|-\rangle\nonumber\\
\widetilde{|\Gamma_8,-\frac{1}{2}\rangle}&&=\frac{\sqrt{2}}{\sqrt{3}}(1-\eta)|iZ\rangle|-\rangle\nonumber\\
&&+\frac{1}{\sqrt{6}}(1+2\eta)[|iX\rangle-i|iY\rangle]|+\rangle\nonumber\\
\end{eqnarray}

The SO contribution in the LH state is plotted as a function of the aspect ratio in Fig.~\ref{fig4}a. The solid line displays the effect of the axial strain, assuming that there is an initial contribution with a high-order envelope function, of amplitude $\eta'$ (that we keep independent of the aspect ratio for simplicity), and the contribution induced by the uniform axial strain, with an s-like envelope of amplitude $\sqrt{2}\eta$ with $\eta=Q/\Delta_{SO}$. As the two envelope functions are orthogonal, we plot $|\eta'|^2+2|\eta|^2$, with $Q$ evaluated using the Eshelby model (Fig.~\ref{fig1}). Figure~\ref{fig4}a shows a good agreement with the numerical data. This approach will be used to evaluate the oscillator strengths in Section \ref{LHSOOS}.

The spin Zeeman Hamiltonian $\mathcal{H}_{sZ}$, Eq.~\ref{Sx} and Eq.~\ref{Sz}, also has non-vanishing matrix elements between the LH and SO states. Simple cases (well isolated LH states, Section~\ref{LHSOSpin}) can be treated in a straightforward manner. With a longitudinal spin Zeeman effect (magnetic field applied along $z$), Eq.~\ref{Sz} and \ref{HQ} allow us to isolate two independent, $2\times2$ matrices:
\begin{eqnarray}\label{LHSOmatrixp}
&&\mathcal{H}_Q+\mathcal{H}_{sZ}+\mathcal{H}_{SO}=\nonumber\\
&&\begin{pmatrix}
-Q+\frac{1}{3}E_{sZ} & -\sqrt{2}Q +\frac{2\sqrt{2}}{3}E_{sZ}\\
-\sqrt{2}Q +\frac{2\sqrt{2}}{3}E_{sZ}& -\Delta_{SO}-\frac{1}{3}E_{sZ}
\end{pmatrix}
\end{eqnarray}
and
\begin{equation}\label{LHSOmatrixm}
\begin{pmatrix}
-Q-\frac{1}{3}E_{sZ} & \sqrt{2}Q +\frac{2\sqrt{2}}{3}E_{sZ}\\
\sqrt{2}Q +\frac{2\sqrt{2}}{3}E_{sZ}& -\Delta_{SO}+\frac{1}{3}E_{sZ}
\end{pmatrix}
\end{equation}
in the $\left(|\Gamma_8, \frac{1}{2}\rangle\oplus |\Gamma_7, \frac{1}{2}\rangle\right)$ and $\left(|\Gamma_8, -\frac{1}{2}\rangle\oplus |\Gamma_7, -\frac{1}{2}\rangle\right)$ doublets, respectively. When the field is applied along $x$, $2\times2$ matrices exhibiting the same structures are obtained using Eqs.~\ref{Hxsym} and \ref{Hxanti} of Appendix~\ref{app:Hamiltonian}:
\begin{eqnarray}\label{100LHSOmatrixp}
&&\mathcal{H}_Q+\mathcal{H}_{sZ}+\mathcal{H}_{SO}=\nonumber\\
&&\begin{pmatrix}
-Q+\frac{2}{3}E_{sZ} & -\sqrt{2}Q +\frac{\sqrt{2}}{3}E_{sZ}\\
-\sqrt{2}Q +\frac{\sqrt{2}}{3}E_{sZ}& -\Delta_{SO}+\frac{1}{3}E_{sZ}
\end{pmatrix}
\end{eqnarray}
and
\begin{equation}\label{100LHSOmatrixm}
\begin{pmatrix}
-Q-\frac{2}{3}E_{sZ} & -\sqrt{2}Q -\frac{\sqrt{2}}{3}E_{sZ}\\
-\sqrt{2}Q - \frac{\sqrt{2}}{3}E_{sZ}& -\Delta_{SO}-\frac{1}{3}E_{sZ}
\end{pmatrix}.
\end{equation}

$\mathcal{H}_Q+\mathcal{H}_{SO}$ represent a balance, within the LH-SO quadruplet, between spin-orbit and crystal field effects (the axial strain in zinc-blende, or a wurtzite structure). As a result, in the axially strained zinc-blende structure, the LH states incorporate a SO contribution, obtained by diagonalizing the above two matrices, and tend towards the $|iZ\rangle|\pm\rangle$ doublet when $(-Q/\Delta_{SO})>>1$.

In order to address more complex cases, for instance the detailed fits of the LH-HH and LH-LH anticrossings in Section~\ref{Field}, it is convenient to build a matrix representation of the axial strain + spin Zeeman Hamiltonian restricted to the $\Gamma_8$ (HH and LH) quadruplet. Exploiting the symmetry properties of the system, the general expression is
\begin{eqnarray}\label{eqHspin}
&&\mathcal{H}_{\Gamma_8}=Q
\begin{pmatrix}
1&0 & 0&0 \\
0&-1&0&0 \\
0&0 &-1\\
0& 0&0 &1
\end{pmatrix}\nonumber\\
&&+2E_{sZ}b_z
\begin{pmatrix}
-\frac{1}{2} \alpha_{SO}&0 & 0&0 \\
0&-\frac{1}{6} \beta_{SO} &0&0 \\
0&0 & +\frac{1}{6} \beta_{SO}& 0\\
0& 0&0 &\frac{1}{2} \alpha_{SO}
\end{pmatrix}\nonumber\\
&& +2E_{sZ}b_x
\begin{pmatrix}
0&\frac{1}{2\sqrt{3}} \gamma_{SO}  &0 &0 \\
\frac{1}{2\sqrt{3}} \gamma_{SO} &0&\frac{1}{3} \delta_{SO} & \\
0&\frac{1}{3} \delta_{SO}  & 0&\frac{1}{2\sqrt{3}} \gamma_{SO}  \\
0&0 &\frac{1}{2\sqrt{3}} \gamma_{SO}  &0
\end{pmatrix}
\end{eqnarray}

with "correction factors" $\alpha_{SO}$,... $\delta_{SO}$ whose deviations from unity represent the effect of the axial-strain coupling with the SO doublet \footnote{These factors may also be used to account for any additional correction due to a mechanism with the same symmetry, or a higher symmetry: for instance, in a nanostructure, a contribution from the Luttinger Hamiltonian, see Section~\ref{Conclusion}, or the probability of presence within the magnetic QD as treated in Section~\ref{Confinement}.}.

Usual techniques (actually, calculating the matrix elements between the modified states, see Eq.~\ref{MixedStates} \cite{AbragamBleaney}) allow us to obtain to first order in $\eta$:
\begin{eqnarray}\label{eq9}
&&\alpha_{SO}=1\nonumber\\
&&\beta_{SO}=1-8\eta\nonumber\\
&&\gamma_{SO}=1+2\eta\nonumber\\
&&\delta_{SO}=1-2\eta\nonumber\\
\end{eqnarray} with $\eta=\frac{Q}{\Delta_{SO}}$.

\subsection{Oscillator strengths} \label{LHSOOS}

Here, we calculate the effect of the mixing of the light hole with the split-off states on the oscillator strength. For the sake of simplicity, we do this calculation in the first order approximation using Eq.~\ref{MixedStates}. The LH and SO components feature similar, well-confined s-like envelope functions, pushed along $z$ by the axial piezoelectric field (see Fig.~\ref{fig2}c). The electron envelope function is also s-like and pushed in the opposite direction (not shown in Fig.~\ref{fig2}c). The overlap between the envelope functions affects the absolute values of the oscillator strengths independently of the orientation of the electron-hole dipole, so that we can ignore its effect when plotting the oscillator-strength ratio. Here also, the point is that the electron-hole dipole matrix element depends on the \emph{amplitude} $\eta$ rather than on the weight $|\eta|^2$: Using Eq.~\ref{MixedStates} and \ref{dipole}, we obtain to first order in $\eta=\frac{Q}{\Delta_0}$
\begin{eqnarray}\label{OSRatio}
&&P_x^2+P_y^2=\frac{1}{3}(1+4\eta)\nonumber\\
&&P_z^2=\frac{2}{3}(1-2\eta)
\end{eqnarray}
The comparison with the numerical calculation in Fig.~\ref{fig4}b shows that the axial-strain mixing of bulk CdTe captures the essentials of the mechanism that determines the evolution of the oscillator strengths. This evolution results from the enhancement of the $|iZ\rangle$ component of the light-hole wavefunction in Eq.~\ref{MixedStates}, which tends to concentrate the whole oscillator strength into a single dipole orientation.

\subsection{Spin properties} \label{LHSOSpin}

The Zeeman shift of an otherwise well-isolated light-hole state coupled to split-off states is addressed by diagonalizing the $2\times2$ matrices, Eq.~\ref{LHSOmatrixp} to \ref{100LHSOmatrixm}. Figure~\ref{fig8} shows that both the SO state's weight and its dependence on the spin Zeeman effect, as well as the resulting Zeeman shifts, are convincingly explained. We took into account the calculated probability to be a LH in the QD, which was 0.93. Then the only adjustable parameter is the axial-strain parameter, set at $Q$=-86~meV. This is not far from the average value, $Q$=-77~meV, calculated numerically in this QD with a (Zn,Mg)Te external shell which slightly reduces the strain built in the dot. Note that our estimate of the spin enhancement by the axial strain is essentially justified for the bulk material: In the case of a QD, we may expect an additional contribution from the confinement, through matrix elements of the Luttinger Hamiltonian between LH and SO states. In the present case, as discussed in Section~\ref{Conclusion}, this contribution is much smaller than the effect of strain.

In Fig.~\ref{fig8}, we may notice local deviations when the spin-down state of the ground Kramers doublet becomes degenerate with the excited states (see the right half of Fig.~\ref{fig5}b, at $E_{sZ}=17$ and 35~meV): Barely visible in the energy (Fig.~\ref{fig8}a), these deviations appear more distinctively in the probability of presence (Fig.\ref{fig8}b). The deviations suggest the presence of a very weak interaction between the two sets of states, leading to anticrossings. More sizable interactions have been addressed in Section \ref{LHLH}.

\subsection{Electron-hole exchange} \label{LHSOeh}

Finally, we consider the effect of the mixing of the light hole with the split-off states on the electron-hole exchange interaction in a quantum dot.

It has been known for some time now that the fine-structure splitting of the heavy-hole exciton can be changed, and even made to vanish, by applying an in-plane stress \cite{Yang2020}. A model evaluating the effect of the Bir and Pikus Hamiltonian on the electron-HH exchange has been proposed in Ref.~\cite{Wang2014}.

\begin{figure} []
\centering
\includegraphics [width=\columnwidth]{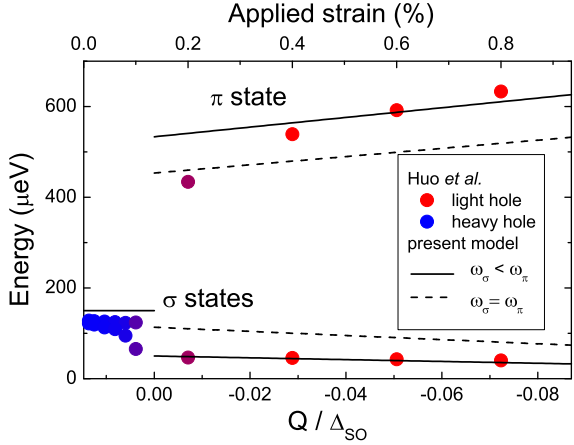}
\caption{Splitting of the exciton's fine structure \emph{vs.} strain. Symbols show the numerically calculated energy \cite{Huo2014} of the $\sigma$-emitting and the $\pi$-emitting exciton states (polarized along $xy$ and along $z$, respectively) in a flat GaAs QD under in-plane tensile strain (top scale). The transition from HHs (blue symbols) to LHs (red symbols) is not abrupt. The lines show the present calculation of the effect of LH-SO mixing, as a function of the strain shift / spin-orbit coupling ratio (bottom scale). The dashed lines use $\omega_\sigma=\omega_\pi=320~\mu$eV, the solid lines $\omega_\sigma=150~\mu$eV and $\omega_\pi=~400 \mu$eV for the LH exciton, and $\omega_\sigma=150~\mu$eV for the HH exciton. The origin of the bottom scale is positioned at the value of the applied strain (top scale) which induces the switching from HH to LH. } \label{fig9}
\end{figure}

A recent example, particularly interesting in the present context, is the case of the flat GaAs quantum dot submitted to an in-plane tensile strain \cite{Huo2014}. The fine structure calculated numerically in Ref.~\cite{Huo2014} displays two characteristic features (Fig.~\ref{fig9}): (1) an increase of the splitting at the HH-LH crossing, from a small splitting between the dark and bright states of the HH exciton, to a 5 times larger splitting between the dark and the \emph{z}-polarized state of the LH exciton ($\pi$-state in Fig.~\ref{fig9}); (2) a further increase of the splitting upon increasing the strain. This increase was confirmed experimentally in the same study \cite{Huo2014}. In addition to the numerical calculation, the electron-hole exchange was discussed using symmetry arguments (theory of invariants \cite{BirPikus}, or the spin Hamiltonian technique \cite{AbragamBleaney}). This approach is currently used to describe the exchange interaction between electrons and HHs \cite{Kesteren1990,Bayer2002}, with a small number of fitting parameters. However, the extension of this approach to light holes requires the introduction of another set of parameters, so that the origin of the two features revealed in Ref.~ \cite{Huo2014} could not be identified precisely. We show now that the jump of the fine-structure splitting at the HH-LH crossing can be ascribed to the flat shape of the QD, and its behavior above the crossing can be ascribed to the LH-SO mixing.

In Appendix~\ref{app:ehexchange}, we extend the description of the HH exciton fine structure, proposed in Ref.~\cite{Kadantsev2010}, to the LH and SO bands. Using Eq.~\ref{ehLH}, which describes the fine structure of the pure LH exciton with two parameters, $\omega_\pi$ and $\omega_\sigma$, and Eq.~\ref{MixedStates}, we obtain the following Hamiltonian for exciton states formed on the SO-mixed LH states:
\begin{equation} \label{ehmixed}
\begin{pmatrix}
|+0\rangle& |+1\rangle& |-1\rangle & |-0\rangle\\
\hline
\frac{2\omega_\pi }{3} (1-2\eta) & 0 & 0 & \frac{2\omega_\pi }{3} (1-2\eta) \\
0 & \frac{\omega_\sigma}{3}  (1+4\eta) & 0 & 0 \\
0 & 0 & \frac{\omega_\sigma}{3}  (1+4\eta) & 0 \\
\frac{2\omega_\pi}{3}  (1-2\eta) & 0 & 0 & \frac{2\omega_\pi }{3} (1-2\eta).
\end{pmatrix}
\end{equation}
The light-hole exciton states are denoted $|+0\rangle$ (transition from $\widetilde{|\frac{1}{2}\rangle}$ in the valence band to $|\frac{1}{2}\rangle$ in the conduction band), $|+1\rangle$ (transition from $\widetilde{|\frac{-1}{2}\rangle}$ to $|\frac{1}{2}\rangle$),  $|-1\rangle$ (transition from $\widetilde{|\frac{1}{2}\rangle}$ to $|\frac{-1}{2}\rangle$) and $|-0\rangle$ (transition from $\widetilde{|\frac{-1}{2}\rangle}$ to $|\frac{-1}{2}\rangle$).
Thus the eigenenergies are now 0 (dark state), $\frac{\omega_\sigma}{3}(1+4\eta)$ and $\frac{4\omega_\pi}{3}(1-2\eta)$.

The parameters $\omega_\sigma$ and $\omega_\pi$ are expected to depend on the QD's geometry, and especially (see Appendix~\ref{app:ehexchange}) on the diameter $D$ and length $L$ of the QD. These parameters are fixed in the case where a stress is applied to a QD as in the work reported in Ref.~\cite{Huo2014}, which describes the shift of the LH fine-structure levels when the stress is increased above the HH-LH crossing. Figure~\ref{fig9} compares the numerical data of Ref.~\cite{Huo2014} (symbols), and the present bulk-type description (lines). The splitting due to the axial strain is calculated as $2Q=-b\left[\varepsilon_{zz}-\frac{1}{2} (\varepsilon_{xx}+\varepsilon_{yy})\right]$, with $\varepsilon_{xx}=\varepsilon_{yy}=f$, the applied strain, and $\varepsilon_{zz}=-\frac{2c_{12}}{c_{11}}f$. We use the GaAs parameters given in Appendix \ref{app:MatPara} to calculate $Q/\Delta_0$ (the bottom scale of Fig.~\ref{fig9}). The origin of this scale is positioned at the HH-LH crossing: In such a flat QD, the effect of the strain has to counterbalance the effect of confinement. The transition is not abrupt, and actually the LH-ground state regime is fully achieved only for the last three points in Fig.~\ref{fig9}. The two adjustable parameters are $\omega_\sigma$ and $\omega_\pi$, set to 150 and 400 $\mu$eV respectively, in agreement with the flat shape of the QD for which we expect $\omega_\sigma<\omega_\pi$ (see Appendix~\ref{app:ehexchange}). The dashed lines are obtained with a common value (as assumed in the description of the electron-hole exchange in nanocrystallites \cite{Sinito2014}), $\omega_\sigma=\omega_\pi=320~\mu$eV. In both cases the effect of the strain is correctly described, which supports strain-induced LH-SO mixing as the driving mechanism for the evolution of the fine structure splitting of the light-hole exciton upon increasing the strain \footnote{The evolution of the fine structure splitting upon increasing the $L/D$ aspect ratio of a compressively-strained QD involves two mechanisms which act in opposite directions: The larger aspect ratio directly decreases the electron-hole exchange energy $\omega_\pi$, see Appendix \ref{app:ehexchange}, but also increases the axial strain, which in turn increases $(1+4\eta)$. The net result in Ref.~\cite{Zielinski2013} is a decrease of the splitting.}.

\subsection{Discussion on LH-SO mixing} \label{}

To sum up, the LH-SO mixing by the axial strain, although small (SO weights around a few $\%$), does induce strong modifications of the oscillator strength, and of the spin Zeeman effect, and of the electron-LH exchange interaction \footnote{Note that we do not expect a similar effect from the conduction band: The uniform strain Hamiltonian, which plays the main role here, does not couple the conduction and valence band states since they have a different parity.}. These modifications account for the main trends of the properties of the LH confined in an elongated quantum dot with compressive lattice-mismatch, or a flat dot submitted to an in-plane tensile strain ($Q<0$). Of course, the effect is reversed for the excited light-hole states in a dot with $Q>0$, such that the ground state is heavy-hole.

We may note that the Luttinger Hamiltonian contains a term $-\gamma_3 \frac{\hbar^2}{2m_0}(\frac{\partial^2}{\partial x^2}+\frac{\partial^2}{\partial y^2}-2\frac{\partial^2}{\partial z^2})$ which has the same symmetry as the coupling to axial strain \cite{Fishman1995}. Its effect should be added to the parameter $Q$ defined previously, and it contributes to the splitting between HH and LH confined states, and to the LH-SO mixing. The two contributions add together in an elongated QD with compressive strain and they partially cancel in a flat QD with tensile strain. In the present case (strong mismatch, several nm dot size), a closer examination of the matrix elements calculated numerically shows that the effect of the axial strain is definitely stronger than the Luttinger term. That may not be true in a QD of smaller size or with a smaller mismatch.

\begin{figure} []
\centering
\includegraphics [width=\columnwidth]{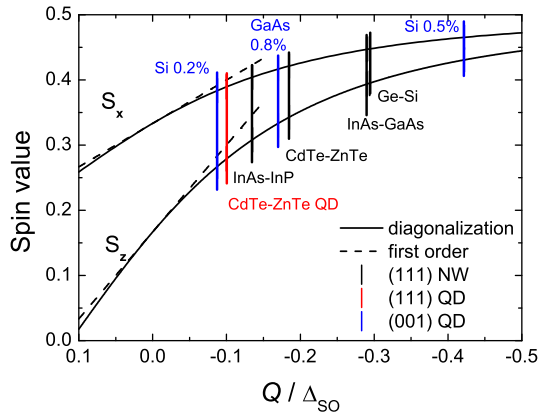}
\caption{Spin values calculated from Eq.~\ref{fullspin} (solid lines) and the linear approximation (Eq.~\ref{eq9}, dashed lines); Different combinations of materials are reported: GaAs and Si QDs on (001) substrates with applied in-plane tensile strain as indicated; CdTe-ZnTe elongated QD with $L/D$=2.25; four different (111) core-shell nanowires.} \label{fig10}
\end{figure}

We can calculate the effective spin along $x$ or $z$, by developing the eigenenergies of the $2\times2$ matrices, Eqs.~\ref{LHSOmatrixp} to \ref{100LHSOmatrixm}, to first order in $E_{sZ}$. The result,
\begin{eqnarray}\label{fullspin}
\langle S_x \rangle =\frac{1}{4}+\frac{1}{12}\frac{\Delta_{SO}-9Q}{\sqrt{(\Delta_{SO}-Q)^2+8Q^2}} \nonumber\\
\langle S_z \rangle =\frac{1}{6}\frac{\Delta_{SO}-9Q}{\sqrt{(\Delta_{SO}-Q)^2+8Q^2}}
\end{eqnarray}
is shown as a function of $\frac{Q}{\Delta_{SO}}$ by the solid lines in Fig.~\ref{fig10}.

Also shown is the ratio of the strain splitting to spin-orbit splitting expected for several materials and nanostructures. The two previous cases, the elongated CdTe QD in ZnTe, and the GaAs flat QD under applied strain, appear quite similar.

Strained silicon structures can reach very large values of $Q/\Delta_{SO}$ owing to the low spin-orbit coupling in this material: This is particularly important because such structures are currently studied for the realization of spin qubits \cite{Venitucci2018}. Figure~\ref{fig10} shows that a Si structure with a residual strain of $0.2\%$ (a value that may be reached in CMOS structures \cite{Venitucci2018}) makes $Q/\Delta_{SO}$ already much larger than in our elongated CdTe-ZnTe QD. Extreme values could be reached by applying an extrinsic strain as done in GaAs in Ref.~\cite{Huo2014}. This may open new opportunities for the control of spin-orbit coupling in hole qubits \cite{Venitucci2018}. It is likely however that a non-perturbative 6-bands \textbf{k.p} model is needed at such large values of $Q/\Delta_{SO}$, as done in the case of strained wide bandgap semiconductors with the wurtzite structure \cite{Gil1997} where a modification of the oscillator strengths by an axial strain has been described.

Other structures in Fig.~\ref{fig10} are core-shell nanowires with various combinations of materials. To calculate the built-in strain, we used the method of Ref.~\cite{Ferrand2014}, in a simplified version where we assume that the shell has the same elastic coefficients as the core. The elongated QD of the present study is still far from the limit of the core-shell CdTe-ZnTe nanowire. The position of the InAs-InP nanowire is due to a small value of the strain splitting which overcompensates the decrease of the spin-orbit interaction. The InAs-GaAs nanowire benefits from the larger lattice mismatch. Finally, the Ge-Si nanowire is characterized by a smaller mismatch but also a smaller spin-orbit coupling. We note that a similar strain effect was invoked in Si-Ge nanostructures \cite{Bottegoni2012} where the LH-HH coupling was shown to impact the orbital part of the Zeeman effect.

\section{Discussion and conclusion} \label{Conclusion}
In this study, we have developed a numerical calculation of the properties of holes localized in lattice-mismatched quantum dots, and we propose and test analytical, phenomenological, and predictive models of these properties.

One goal of this study was to test spin Hamiltonians for a valence hole in a QD containing magnetic impurities. In a bulk dilute magnetic semiconductor, the interaction of the $\Gamma_8$ electrons with the ensemble of magnetic impurities is described by a spin Zeeman effect proportional to their magnetization, which splits the quadruplet - the so-called giant Zeeman effect. This approach is also correct in the presence of a strain, or in a quantum well, two cases where the quadruplet is split into a HH doublet and a LH doublet. This quadruplet model was further used to describe QDs containing a dilute magnetic semiconductor, in spite of the large number of excited states which are present between the ground state and the first state of opposite type, and are coupled to the ground state.

Most often, the ground state is the heavy hole, and the most relevant fingerprint of this is the shift induced by a transverse magnetic field, usually interpreted as the presence of the anticrossing between the HH ground state and a LH state of similar envelope function, separated by $\sim 2Q$. The present analysis of Fig.~\ref{fig5}a and Fig.~\ref{fig6}b shows that this is a reasonable description although the anticrossing is probably not related to a single, well identified LH partner, particularly if the valence band offset is small. Such a behavior was observed experimentally in a (Zn,Mn)Te / (Zn,Mg)Te core-shell nanowire \cite{Szymura2015} and the parameters of the quadruplet model (with $2Q$=50~meV) were found to be in good agreement with the tensile strain expected from the structure geometry. In the same study \cite{Szymura2015}, a HH ground state was also found in a (Cd,Mn)Te QD, and features the same characteristic Zeeman effect under a transverse field. As the strain in the dot is compressive, this suggests a flat dot configuration, but the presence of the external (Zn,Mg)Te shell extends the HH ground state domain so that it can include slightly elongated dots \cite{Artioli2019}.

A LH ground state was observed in compressively-strained elongated (Cd,Mn)Te QDs in ZnTe \cite{Jeannin2017} and in (Cd,Mn)Te / (Cd,Mg)Te core-shell nanowires \cite{Plachka2018}. In the latter case, the anisotropy of the Zeeman effect was measured and found to be in agreement with the $\Gamma_8$ model, with $2Q=-10$~meV determined from the anticrossing induced by a transverse field. This small value is consistent with the strain expected in this structure. It is nonetheless too small to reveal the mixing with the SO states (calculated $\eta<10^{-3}$). This coupling should be looked for in structures with a larger mismatch, such as elongated (Cd,Mn)Te QDs in ZnTe. Then the giant Zeeman shift should be described by taking into account the spin enhancement (Eq.~\ref{eqHspin}) but also the quadratic terms due to anticrossing with other LH states (Section~\ref{LHLH}) and the change of confinement (Section~\ref{Confinement}).

Our results have a much broader impact, well beyond the domain of dilute magnetic semiconductor nanostructures.

The inclusion of a spin Zeeman effect of large amplitude allows us to reveal anticrossings between the ground state and various states present between the ground state and the first state of opposite type. The role of such states has been pointed out recently for (001) InAs-GaAs QDs \cite{Luo2015}. Our system has a particularly high symmetry: It is (111)-oriented in the zinc-blende structure, so that several mechanisms present in (001) QDs are eliminated \cite{Singh2009, Schliwa2009, Ha2019}, and its shape is an ellipsoid of revolution. We thus ascribe possible mixing effects to the non-homogeneous components of the built-in strain, which are significant essentially in the shell around the QD, and leak slightly into it (Fig.~\ref{fig1}). As a result, anticrossing takes place between the ground LH state and the excited LH states in the shallow QD (Fig.~\ref{fig5}d) while they are barely visible in the deep QD (Fig.~\ref{fig5}c).

An important result is the strong effect of the coupling between the LH and SO states induced by the axial strain. It appears as a very general mechanism which deeply affects the most important characteristics of the LH state. This mechanism has been known for a long time, but considered as marginal \cite{Fishman2010} because of the modest amplitude of the strain in bulk materials or quantum wells, added to the fact that it does not affect the HH states which form the ground state in quantum wells. Its role was underlined for materials with a small spin-orbit coupling, such as GaN \cite{Gil1997}. Built-in strain can be quite high in quantum dots such as CdTe dots in ZnTe or InAs dots in GaAs. We have shown that the key parameter is indeed the ratio $\eta=Q/\Delta_0$ of the valence band shift (due to the axial strain) to the spin-orbit coupling: Current structures display values of $\mid \eta\mid$ ranging from 0.1 to 0.5, due either to the mismatch strain in core-shell or QD structures, or to a stress applied to the nanostructure, or even to the residual strain in a Si nanodevice. We have shown that:

\begin{itemize}
 \item  The oscillator strength of the LH exciton is strongly affected, enhancing the dipole oriented along the strain axis ($\pi$-polarized emission). For practical purposes, this has to be combined with the effect of dielectric screening and guiding; It is expected to affect the emission of classical light as well as single-photons, and the optical manipulation of qubits in III-V or II-VI nanostructures.
 \item The spin of the LH is strongly modified and pushed towards an isotropic $\frac{1}{2}$ spin when the ground state is LH (and away from $\frac{1}{2}$ if the ground state is HH). Beyond the exchange with a magnetic impurity, further calculations are needed to address the Zeeman effect in a dot which does not contain a dilute magnetic semiconductor, and the response to an applied microwave, for instance aiming at qubit manipulation. As the spin-orbit coupling is particularly small in silicon, this is of special interest for the silicon nanodevices being developed at present for quantum information processing.
 \item  We have extended to the light-hole exciton the treatment of electron-hole exchange previously developed for heavy-hole excitons. It allows us to satisfactorily describe the jump in the fine structure when a stress is applied to a quantum dot to switch the ground state from HH to LH, and to describe the further shift of the LH exciton lines when the stress is increased above the HH-LH crossing. The splitting between the LH $\pi$-emitting bright state and the dark state is larger than the splitting between the $\sigma$-emitting bright states and the dark state. The ratio of these two splittings is equal to 4 if only short-range exchange is taken into account. When long range exchange is included, the ratio tends to be larger than 4 in a flat dot and smaller than 4 in an elongated one. The ratio is also modified by the built-in axial strain, or by an applied stress. Such a manipulation of the fine structure of the light-hole exciton in an elongated quantum dot offers an opportunity to tune the splitting between the $\sigma$ and $\pi$-polarized optical transitions.
\end{itemize}

\begin{acknowledgments}
We acknowledge funding by the French National Research Agency (project ESPADON
ANR-15-CE24-0029). We thank Ronald Cox for his invaluable critical reading of the manuscript.
\end{acknowledgments}

\appendix

\section{Hamiltonian} \label{app:Hamiltonian}

The Hamiltonian is written for the electron states in the $\Gamma_8\oplus \Gamma_7$ multiplet.

We define the electron states \cite{Fishman1995, Fishman2010}
\begin{eqnarray}\label{States}
&& |\Gamma_8,\frac{3}{2}\rangle=-\frac{1}{\sqrt{2}}[|iX\rangle+i|iY\rangle]|+\rangle\nonumber\\
&& |\Gamma_8,\frac{1}{2}\rangle=\frac{\sqrt{2}}{\sqrt{3}}|iZ\rangle|+\rangle-\frac{1}{\sqrt{6}}[|iX\rangle+i|iY\rangle]|-\rangle\nonumber\\
&& |\Gamma_8,-\frac{1}{2}\rangle=\frac{\sqrt{2}}{\sqrt{3}}|iZ\rangle|-\rangle+\frac{1}{\sqrt{6}}[|iX\rangle-i|iY\rangle]|+\rangle\nonumber\\
&& |\Gamma_8,-\frac{3}{2}\rangle=\frac{1}{\sqrt{2}}[|iX\rangle-i|iY\rangle]|-\rangle\nonumber\\\nonumber\\
&& |\Gamma_7,\frac{1}{2}\rangle=\frac{1}{\sqrt{3}}|iZ\rangle|+\rangle+\frac{1}{\sqrt{3}}[|iX\rangle+i|iY\rangle]|-\rangle\nonumber\\
&& |\Gamma_7,-\frac{1}{2}\rangle=-\frac{1}{\sqrt{3}}|iZ\rangle|-\rangle+\frac{1}{\sqrt{3}}[|iX\rangle-i|iY\rangle]|+\rangle\nonumber\\
\end{eqnarray}

We also use symmetric and antisymmetric superpositions of these states, for instance
\begin{eqnarray}\label{SymAnti}
&& |HH_{sym}\rangle=\frac{1}{\sqrt{2}}\left[|\Gamma_8,\frac{3}{2}\rangle+|\Gamma_8,-\frac{3}{2}\rangle \right],\nonumber\\
&& |HH_{anti}\rangle=\frac{1}{\sqrt{2}}\left[|\Gamma_8,\frac{3}{2}\rangle-|\Gamma_8,-\frac{3}{2}\rangle \right],
\end{eqnarray} and so on.

The Luttinger Hamiltonian and its expression with different quantization axes are given in Ref. \cite{Fishman1995, Fishman2010}.

The axial strain Hamiltonian is
\begin{equation} \label{HQ}
\mathcal{H}_Q = Q
\begin{pmatrix}
1 & 0 & 0 & 0 & 0 & 0\\
0 & -1 & 0 & 0 & - \sqrt{2} & 0\\
0 & 0 & -1 & 0 & 0 & \sqrt{2}\\
0 & 0 & 0 & 1 & 0 & 0\\
0 & -\sqrt{2} & 0 & 0 & 0 & 0\\
0 & 0 & \sqrt{2} & 0 & 0 & 0
\end{pmatrix}
\end{equation}
with $Q=-b\left[\varepsilon_{zz}-\frac{1}{2} (\varepsilon_{xx}+\varepsilon_{yy})\right]$ if the symmetry axis is $\langle001\rangle$ and $Q=-\frac{d}{\sqrt{3}}\left[\varepsilon_{zz}-\frac{1}{2} (\varepsilon_{xx}+\varepsilon_{yy})\right]$ if the symmetry axis is $\langle111\rangle$ \cite{Fishman1995}. The sign in $\mathcal{H}_Q$ applies for the electron Hamiltonian, with the convention $b$ and $d<0$, so that a strain with $\left[\varepsilon_{zz}-\frac{1}{2} (\varepsilon_{xx}+\varepsilon_{yy})\right]<0$ pushes the LH valence band up into the gap. Other terms of the Bir and Pikus Hamiltonian (usually named $R$ and $S$ \cite{BirPikus, Fishman2010}) describe the effect of other strain components; They reach significant values essentially in the shell, and they are taken into account only in the numerical calculations.

The spin Zeeman Hamiltonian is written $\mathcal{H}_{sZ}=2E_{sZ}~\textbf{b}.\textbf{S}$, with the following spin matrices within the $\Gamma_8\oplus \Gamma_7$ multiplet:

\begin{equation} \label{Sx}
S_x =
\begin{pmatrix}
0 & \frac{\sqrt{3}}{6} & 0 & 0 & -\frac{1}{\sqrt{6}} & 0\\
\frac{\sqrt{3}}{6} & 0 & \frac{1}{3} & 0 & 0 & -\frac{\sqrt{2}}{6}\\
0 & \frac{1}{3} & 0 & \frac{\sqrt{3}}{6} & \frac{\sqrt{2}}{6} & 0\\
0 & 0 & \frac{\sqrt{3}}{6} & 0 & 0 & \frac{1}{\sqrt{6}}\\
-\frac{1}{\sqrt{6}} & 0 & \frac{\sqrt{2}}{6} & 0 & 0 & -\frac{1}{6}\\
0 & -\frac{\sqrt{2}}{6} & 0 & \frac{1}{\sqrt{6}} & -\frac{1}{6} & 0
\end{pmatrix}
\end{equation}

\begin{equation} \label{Sz}
S_z =
\begin{pmatrix}
\frac{1}{2} & 0 & 0 & 0 & 0 & 0\\
0 & \frac{1}{6} & 0 & 0 &\frac{\sqrt{2}}{3} & 0 \\
0 & 0 & -\frac{1}{6} & 0 & 0 & \frac{\sqrt{2}}{3}\\
0 & 0 & 0 & -\frac{1}{2} & 0 & 0\\
0 & \frac{\sqrt{2}}{3} & 0 & 0 & - \frac{1}{6} & 0\\
0 & 0 & \frac{\sqrt{2}}{3} & 0 & 0 &  \frac{1}{6}
\end{pmatrix}
\end{equation}

Note the similarity between $\mathcal{H}_Q$ and $S_z$, with however a (crucial) change of sign in off-diagonal terms.

When the field is applied along the $x$ axis, a well-adapted basis is formed by the symmetric and antisymmetric superpositions of the $z$-oriented HH $|\Gamma_8,\pm\frac{3}{2}\rangle$,  LH $|\Gamma_8,\pm\frac{1}{2}\rangle$, and SO $|\Gamma_7,\pm\frac{1}{2}\rangle$ states. In the HH$_{sym}$, LH$_{sym}$, SO$_{anti}$ subspace, the matrix representations are
\begin{equation} \label{Hxsym}
S_x =
\begin{pmatrix}
0 & \frac{\sqrt{3}}{6} & -\frac{1}{\sqrt{6}} \\
\frac{\sqrt{3}}{6} & \frac{1}{3} & \frac{\sqrt{2}}{6}\\
-\frac{1}{\sqrt{6}} & \frac{\sqrt{2}}{6} & \frac{1}{6}
\end{pmatrix}
,~
\mathcal{H}_Q = Q
\begin{pmatrix}
1 & 0 & 0 \\
0 & -1 & -\sqrt{2}\\
0 & -\sqrt{2} & 0
\end{pmatrix}
\end{equation}
while in the decoupled, HH$_{anti}$, LH$_{anti}$, SO$_{sym}$ subspace
\begin{equation} \label{Hxanti}
S_x =
\begin{pmatrix}
0 & \frac{\sqrt{3}}{6} & -\frac{1}{\sqrt{6}} \\
\frac{\sqrt{3}}{6} & -\frac{1}{3} & -\frac{\sqrt{2}}{6}\\
-\frac{1}{\sqrt{6}} & -\frac{\sqrt{2}}{6} & -\frac{1}{6}
\end{pmatrix}
,~
\mathcal{H}_Q = Q
\begin{pmatrix}
1 & 0 & 0 \\
0 & -1 & -\sqrt{2}\\
0 & -\sqrt{2} & 0
\end{pmatrix}
\end{equation}
so that the sub-blocks of the Hamiltonian for the LH-SO states are those in Eqs.~\ref{100LHSOmatrixp} and \ref{100LHSOmatrixm}, respectively.

The electric dipole formed by a hole in the valence band and an electron in the conduction band is obtained from the matrix elements
\begin{equation} \label{dipole}
\langle s|p_x|iX \rangle=\langle s|p_y|iY \rangle=\langle s|p_z|iZ \rangle=\varpi.
\end{equation}

\section{Details of the fits} \label{app:fits}

Here we give the details of the fits of Fig.~\ref{fig6}b and Fig.~\ref{fig7}b, which involve several mechanisms.

The built-in axial strain in the flat QD of Fig.~\ref{fig6}b is such that $Q=100$~meV, hence $\gamma_{SO}=1.22$ and $\delta_{SO}=0.78$ from Eq.~\ref{eq9}. Indeed, Fig.~\ref{fig2}a shows that there is a good LH candidate $\sim105$~meV below the HH ground-state. A fit to Fig.~\ref{fig6}b yields $\gamma_{SO}=1.28$ and $\delta_{SO}=0.3$. The low fitting value for $\delta_{SO}$ suggests that the LH state is not as well confined in the QD. Moreover, the Zeeman shift in Fig.~\ref{fig6}b may also result from the interaction with several excited LH states and should not be taken as the signature of a single state. This is confirmed by the fact that the behavior of the shallow, flat QD (Fig.~\ref{fig5}c) is very similar, although in this case the potential consists in an antidot for the LH states.

A good fit is obtained in Fig.~\ref{fig7}b by using the effective Hamiltonian (Eq.~\ref{eqHspin}) for two interacting LH states: the ground state, and an excited state, reasonably well confined in the QD, about 15~meV from the ground state. This energy was determined from the energies at high values of $E_{sZ}$, on both sides of the anticrossing. In order to take into account the confinement and its dependence on the applied field (Section \ref{Confinement}), the parameters of Eq.~\ref{eq9} were scaled by a coefficient equal to the probability of presence in the QD in the same range of values of $E_{sZ}$. Then the matrix element coupling the two LH states is the only remaining parameter. The quality of the fit was evaluated from the plot of the energy (shown in Fig.~\ref{fig7}b) and the plot of the probabilities of presence over the anticrossing (not shown).

\section{Electron - hole exchange} \label{app:ehexchange}

The so-called short-range exchange (SR) is described by an isotropic Hamiltonian, $\omega (\frac{1}{2}-2\textbf{S}^e.\textbf{S}^h)$ (where $\textbf{S}^e$ is the electron spin and $\textbf{S}^h$ the \emph{hole} spin), acting on the exciton (or the electron-hole) states \cite{Fishman2010}. With pure spins, it splits the singlet state, $\frac{|+\rangle_e|-\rangle_h-|-\rangle_e|+\rangle_h}{\sqrt{2}}$, from the triplet states. One can also use valence-electron states instead of hole states, then the singlet state is $\frac{|+\rangle_C|+\rangle_V+|-\rangle_C|-\rangle_V}{\sqrt{2}}$, as a result of the time-reversal properties of a spin $\frac{1}{2}$, and the Hamiltonian is changed accordingly.

If only the $\Gamma_8$ valence band is considered, the exchange interaction can be written $\omega (\frac{1}{2}-\frac{2}{3}\textbf{S}^e.\textbf{J}^h)$. This is the form which is currently used in small nanocrystallites with the zinc-blende structure, together with the anisotropy of the hole - \emph{i.e}., terms in $(2J_z^{h~2}-J_x^{h~2}-J_y^{h~2})$ and $(J_x^{h~2}-J_y^{h~2})$, describing the shape anisotropy acting through the Luttinger Hamiltonian \cite{Gupalov2000, Sinito2014, Goupalov2006, Sercel2018}.

With the exciton states noted $|+2\rangle$ ($|\frac{3}{2}\rangle_h|\frac{1}{2}\rangle_e$), $|+1\rangle$ ($|\frac{+3}{2}\rangle_h|\frac{-1}{2}\rangle_e$),  $|-1\rangle$ ($|\frac{-3}{2}\rangle_h|\frac{1}{2}\rangle_e$) and $|-2\rangle$ ($|\frac{-3}{2}\rangle_h|\frac{-1}{2}\rangle_e$), the short-range exchange Hamiltonian is for the HHs:
\begin{equation} \label{elh}
\omega_{hh}^{SR}
\begin{pmatrix}
|+2\rangle& |+1\rangle& |-1\rangle & |-2\rangle\\
\hline
0 & 0 & 0 & 0\\
0 & 1 & 0 & 0 \\
0 & 0 & 1 & 0 \\
0 & 0 & 0 & 0
\end{pmatrix}.
\end{equation}
The same matrix applies in the valence-electron conduction-electron notation, with the states respectively $|\frac{-3}{2}\rangle_V|\frac{1}{2}\rangle_C$), $|\frac{-3}{2}\rangle_V|\frac{-1}{2}\rangle_C$,  $|\frac{3}{2}\rangle_V|\frac{1}{2}\rangle_C$ and $|\frac{3}{2}\rangle_V|\frac{-1}{2}\rangle_C$.

For pure LHs, using the electron-hole states ($|+0\rangle=|\frac{-1}{2}\rangle_h|\frac{1}{2}\rangle_e$, $|+1\rangle=|\frac{1}{2}\rangle_h|\frac{1}{2}\rangle_e$,  $|-1\rangle=|\frac{-1}{2}\rangle_h|\frac{-1}{2}\rangle_e$ and $|-0\rangle=|\frac{1}{2}\rangle_h|\frac{-1}{2}\rangle_e$, the matrix is
\begin{equation} \label{elh}
\omega_{lh}^{SR}
\begin{pmatrix}
|+0\rangle& |+1\rangle& |-1\rangle & |-0\rangle\\
\hline
\frac{2}{3} & 0 & 0 & -\frac{2}{3}\\
0 & \frac{1}{3} & 0 & 0 \\
0 & 0 & \frac{1}{3} & 0 \\
-\frac{2}{3} & 0 & 0 & \frac{2}{3}
\end{pmatrix}.
\end{equation}
The same matrix with all terms with positive sign holds when using the valence-electron conduction-electron states.

For the HH exciton, this is the usual result that the two dark states, $|\pm 2\rangle$, remain degenerate and unshifted, while the two bright states, $|\pm1\rangle$, remain degenerate but are shifted by $\omega_{hh}^{SR}$. For the LH exciton, we obtain a dark state $|D\rangle=\frac{1}{\sqrt{2}}(|+0\rangle - |-0\rangle)$ unshifted, a bright doublet $|\pm1\rangle$, emitting $\sigma$-polarized light, upshifted by $\frac{1}{3}\omega_{lh}^{SR}$ and a bright singlet, $|\pi\rangle=\frac{1}{\sqrt{2}}(|+0\rangle + |-0\rangle)$, emitting $\pi$-polarized light, upshifted by $\frac{4}{3}\omega_{lh}^{SR}$. These three levels are those described in Ref.~\cite{Huo2014}. The result applies to pure LH states. \footnote{For SO excitons, due to phase convention, we have $|D\rangle=\frac{1}{\sqrt{2}}(|+0\rangle + |-0\rangle)$ and $|\pi\rangle=\frac{1}{\sqrt{2}}(|+0\rangle - |-0\rangle)$.}

In the case of the HH, it is well-known that the bright doublet splits due to long-range (LR) electron-hole exchange. Experimental data are usually described phenomenologically by a spin Hamiltonian \cite{AbragamBleaney, BirPikus} acting within the electron-HH exciton quadruplet and containing off-diagonal terms \cite{Kesteren1990,Bayer2002}. These terms include contributions from the reduced symmetry of the atomistic potential present in (001)-oriented dots \cite{Bester2003} but not in (111)-oriented dots \cite{Singh2009, Schliwa2009, Ha2019}. Other contributions are due to the reduced (mesoscopic) symmetry of the shape of the dot.

A more complete discussion incorporating these long-range terms, adapted to confined systems, was introduced by Maialle \cite{Maialle1993} for excitons in a quantum well. The electron-hole states are used, and the appropriate distinction is made between the envelope functions of HHs and LHs. The case of quantum dots formed by interface fluctuations in a quantum well was considered by Takagahara \cite{Takagahara2000}; The matrix elements, including the prefactors, are calculated explicitly for the SR contribution: As the basis used is the product of conduction-band electron states and valence-band electron states, the matrix elements are proportional to the overlap of electron and hole Bloch functions with the same spin, multiplied by a weighted overlap of the electron and hole envelope functions. These terms have been detailed in Ref.~\cite{Kadantsev2010}, assuming that the exciton is strongly confined in a QD (weak electron-hole correlations), so that the two-particle wavefunction can be written simply as the product of a single hole and a single electron state. This study was restricted to the HH excitons.

An extension of Ref.~\cite{Kadantsev2010} leads us to define
\begin{equation}\label{delta}
\Delta_{\alpha\beta}= (E^{SR}-\mu^2\frac{8\pi}{3}) R_0 \delta_{\alpha\beta}-2\mu^2R_{\alpha\beta}
\end{equation}
where \cite{Kadantsev2010}
\begin{eqnarray}\label{integrales}
R_0&&=  \int F_c^*(\textbf{r}) F_v(\textbf{r}) F_{c'}(\textbf{r}) F_{v'}^*(\textbf{r}) d\textbf{r}\nonumber\\
R_{\alpha\beta}&&= \int\int \left[\frac{\partial^2 F_c^*(\textbf{r}) F_v(\textbf{r})}{\partial \alpha~\partial \beta}\right]\frac{F_{c'}(\textbf{r}') F_{v'}^*(\textbf{r}')}{|\textbf{r}-\textbf{r}'|} d\textbf{r} d\textbf{r}'\nonumber\\
\end{eqnarray}
Letters $\alpha,\beta$ label the cartesian coordinates $=x,y,z$, and the $F_c$'s and $F_v$'s are the envelope functions of the conduction electrons and valence holes, respectively. Equations \ref{delta} and \ref{integrales} are easily understood if the hole states are pure $|X\rangle$, $|Y\rangle$ or $|Z\rangle$ Bloch function. Then, the $\Delta_{\alpha\beta}$'s are the matrix elements of an operator $\Delta$ acting on the orbital part of electron-hole states described by the envelope function $F_c$ of the electron, the envelope function $F_v$ of the hole, and its Bloch state $|\alpha\rangle$, and another state described by $F_{c'}$, $F_{v'}$ $|\beta \rangle$. Due to the derivatives involved in the definition of $R_{\alpha\beta}$ in Eq.~\ref{integrales}, the $\Delta_{\alpha\beta}$'s form a tensor of rank 2.

The definition of ($\Delta$) allows us to write the electron-hole exchange matrix as the product of an orbital part $\Delta$ and a spin part:
\begin{equation}\label{DeltaSpin}
\Delta~~(\frac{1}{2}-2\textbf{S}^e.\textbf{S}^h).
\end{equation}

When considering only the HH excitons, as in Ref.~\cite{Kadantsev2010}, the restriction $\alpha,\beta=x,y$ is sufficient, and the envelope functions $F_v$ and $F_{v'}$ are those of the HH. The extension to LH and SO excitons requires us to calculate the matrix elements of $\Delta$ between HH, LH and SO Bloch states. The final result will be applied to the HH, LH and SO envelope functions.

It is interesting to first recalculate $\Delta$ in the $|+1\rangle$, $|-1\rangle$, $|0\rangle$ basis of Bloch orbital states (or apply the technique of invariants \cite{BirPikus}). With $|+1\rangle=\frac{-|iX\rangle-i|iY\rangle}{\sqrt{2}}$, $|-1\rangle=\frac{|iX\rangle-i|iY\rangle}{\sqrt{2}}$, $|0\rangle=|iZ\rangle$, we obtain:
\begin{equation}\label{deltaOriented}
\Delta=
\begin{pmatrix}
|+1\rangle& |-1\rangle&|0\rangle  \\
\hline
\delta_0& -\delta_1 &-\delta^\prime_1 \\
-\delta_1^* & \delta_0& \delta_1^{\prime*} \\
-\delta^{\prime*}_1 & \delta^\prime_1 & \delta'_0
\end{pmatrix}.
\end{equation}
with
\begin{eqnarray}\label{delta2}
\delta_0&&= E^{SR} R_0-\mu^2(\frac{8\pi}{3} R_0+R_{xx}+R_{yy})\nonumber\\
\delta'_0&&= E^{SR} R_0-\mu^2(\frac{8\pi}{3} R_0+2R_{zz})\nonumber\\
\delta_1&&= \mu^2(R_{xx}-R_{yy}+2iR_{xy})\nonumber\\
\delta'_1&&= -2\mu^2\frac{R_{xz}-iR_{yz}}{\sqrt{2}}\nonumber\\
\end{eqnarray}
Similarly to the Bir-Pikus and Luttinger Hamiltonians, $\Delta$ contains terms with specific properties of symmetry with respect to the quantization axis, $\delta_0$ and $\delta'_0$ conserving the projection of angular momentum (as the $P$ and $Q$ terms), $\delta'_1$ changing it by 1 (as the $S$-term) and $\delta'_1$ by 2 (as $R$).

We finally obtain the following exchange Hamiltonian, expressed in the electron-hole pair states:
\begin{widetext}
\begin{equation} \label{elh}
\left(
\begin{array} {cccc|cccc|cccc}
HH &HH& HH &HH&LH &LH &  LH&LH &SO &SO & SO &SO  \\
|+1\rangle & |-1\rangle & |+2\rangle & |-2\rangle & |+1\rangle & |-1\rangle & |\pi\rangle & |D\rangle & |+1\rangle & |-1\rangle&  |\pi\rangle & |D\rangle\\
\hline
\delta_0 & \delta_1 & 0 & 0&-\frac{1}{\sqrt{3}}\delta_0 &- \frac{1}{\sqrt{3}}\delta_1 & \frac{2}{\sqrt{3}}\delta'_1 & 0&\frac{\sqrt{2}}{\sqrt{3}}\delta_0 & -\frac{\sqrt{2}}{\sqrt{3}}\delta_1&- \frac{\sqrt{2} }{\sqrt{3}} \delta'_1& 0\\
\delta_1^* & \delta_0 & 0 & 0 &-\frac{1}{\sqrt{3}}\delta^*_1 & -\frac{1}{\sqrt{3}}\delta_0 & \frac{2}{\sqrt{3}}\delta^{'*}_1 & 0& \frac{\sqrt{2}}{\sqrt{3}}\delta^*_1 & -\frac{\sqrt{2}}{\sqrt{3}} \delta_0& -\frac{\sqrt{2}}{\sqrt{3}}\delta^{'*}_1 & 0\\
0 & 0 & 0 & 0 &0 & 0 & 0 & 0&0 & 0 & 0 & 0\\
0 & 0 & 0 & 0&0 & 0 & 0 & 0&0 & 0 & 0 & 0\\
\hline
-\frac{1}{\sqrt{3}}\delta_0& -\frac{1}{\sqrt{3}}\delta_1 & 0 & 0&\frac{1}{3} \delta_0&\frac{1}{3}\delta_1&-\frac{2}{3} \delta'_1& 0&-\frac{\sqrt{2}}{3}\delta_0&\frac{\sqrt{2}}{3}\delta_1 & \frac{\sqrt{2}}{3}\delta'_1 & 0\\
-\frac{1}{\sqrt{3}}\delta_1^* &-\frac{1}{\sqrt{3}}\delta_0 & 0 & 0 &\frac{1}{3}\delta_1^* & \frac{1}{3}\delta_0& -\frac{2}{3}\delta'^*_1 & 0&-\frac{\sqrt{2}}{3}\delta_1^*&\frac{\sqrt{2}}{3}\delta_0 & \frac{\sqrt{2}}{3}\delta'^*_1 & 0\\
\frac{2}{\sqrt{3}} \delta'^*_1&\frac{2}{\sqrt{3}}\delta'_1 & 0 & 0 &-\frac{2 }{3}\delta'^*_1 &-\frac{2}{3}\delta'_1 & \frac{4}{3}\delta'_0 & 0&\frac{2\sqrt{2}}{3}\delta'^*_1 & -\frac{2\sqrt{2}}{3}\delta'_1 & -\frac{2\sqrt{2}}{3} \delta'_0 & 0\\
0 & 0 & 0 & 0&0 & 0 & 0 & 0&0 & 0 & 0 & 0\\
\hline
\frac{\sqrt{2}}{\sqrt{3}}\delta_0 & \frac{\sqrt{2}}{\sqrt{3}} \delta_1& 0 & 0&-\frac{\sqrt{2}}{3}\delta_0 &-\frac{\sqrt{2}}{3} \delta_1& \frac{2\sqrt{2}}{3}\delta'_1 & 0&\frac{2}{3} \delta_0& -\frac{2}{3}\delta_1&-\frac{2}{3} \delta'_1& 0\\
-\frac{\sqrt{2}}{\sqrt{3}} \delta_1^*& -\frac{\sqrt{2}}{\sqrt{3}}\delta_0 & 0 & 0 &\frac{\sqrt{2}}{3}\delta_1^* &\frac{\sqrt{2}}{3}\delta_0 &-\frac{2\sqrt{2}}{3}\delta'^*_1 & 0& -\frac{2}{3}\delta_1^* & \frac{2}{3} \delta_0& \frac{2}{3} \delta'^*_1& 0\\
-\frac{\sqrt{2}}{\sqrt{3}}\delta'^*_1 & -\frac{\sqrt{2}}{\sqrt{3}}\delta'_1  & 0 & 0 &\frac{\sqrt{2}}{3}\delta'^*_1 &\frac{\sqrt{2}}{3} \delta'_1 &-\frac{2\sqrt{2}}{3} \delta'_0&0&-\frac{2}{3}\delta'^*_1  &\frac{2}{3}\delta'_1  & \frac{2}{3}\delta'_0 &0\\
0 & 0 & 0 & 0&0 & 0 & 0 & 0&0 & 0 & 0 & 0\\
\end{array}
\right)
\end{equation}
\end{widetext}

with the $\delta_i$'s given in Eq.~\ref{delta2}.

The matrix is composed of $3\times3$ blocks, formed on Eq.~\ref{deltaOriented}, each of them corresponding to the HH, LH and SO excitons; It is understood that the envelope functions $F_c(\textbf{r})$ and $F_v(\textbf{r})$ (hence the parameters $\delta_i$ as well) are \emph{a priori} different in each block, as in Ref.~\cite{Maialle1993}. $E^{SR}$ and $\mu$ characterize the short-range and long-range exchange, respectively, and are discussed in Ref.~\cite{Kadantsev2010}. The previously defined $\omega^{SR}=E^{SR} R_0$.

The discussion about the $\delta_1$ term which has been developed in Ref.~\cite{Kadantsev2010} for the HH exciton is valid also for the LH and the SO excitons: A proper choice of the $x$ and $y$ axes makes $R_{xy}$ vanish, and if the symmetry within the $xy$ plane is high enough (circular, or square $D_{2d}$, or trigonal $C_{3v}$), $R_{xx}=R_{yy}$ and $\delta_1=0$: Then the bright $\sigma$ doublet remains degenerate. The same argument shows that under such conditions of high symmetry, $R_{zx}$ and $R_{zy}$ also vanish and $\delta'_1=0$. However, the diagonal terms, for instance $\frac{1}{3}\delta_0$ and $\frac{4}{3}\delta'_0$ for the LH exciton, remain distinct.

In a further step proposed in Ref.~\cite{Kadantsev2010}, the $R_{\alpha\beta}$ terms were calculated using Gaussian envelope functions, $\exp (-\alpha_x x^2-\alpha_y y^2 -\alpha_z z^2)$ (harmonic oscillator approximation). The three parameters $\alpha_x$, $\alpha_y$, $\alpha_z$ characterize the extension of the envelope function  along the corresponding direction, and decrease as the envelope function expands. Then $R_{xx}$, $R_{yy}$ and $R_{zz}$ are proportional to
\begin{eqnarray}\label{IxIyIz}
I_x&&=\int_0^{\frac{\pi}{2}} d\phi \int_0^{\frac{\pi}{2}} \sin \theta d\theta \frac{\sin^2 \theta \cos^2 \phi}{\sin^2 \theta (\frac{\sin^2 \phi}{2\alpha_x}+\frac{\cos^2 \phi}{2\alpha_y})+\frac{\cos^2 \theta}{2\alpha_z}}\nonumber\\
I_y&&=\int_0^{\frac{\pi}{2}} d\phi \int_0^{\frac{\pi}{2}} \sin \theta d\theta \frac{\sin^2 \theta \sin^2 \phi}{\sin^2 \theta (\frac{\sin^2 \phi}{2\alpha_x}+\frac{\cos^2 \phi}{2\alpha_y})+\frac{\cos^2 \theta}{2\alpha_z}}\nonumber\\
I_z&&=\int_0^{\frac{\pi}{2}} d\phi \int_0^{\frac{\pi}{2}} \sin \theta d\theta \frac{\cos^2 \theta }{\sin^2 \theta (\frac{\sin^2 \phi}{2\alpha_x}+\frac{\cos^2 \phi}{2\alpha_y})+\frac{\cos^2 \theta}{2\alpha_z}}\nonumber\\
\end{eqnarray}

In Ref.~\cite{Kadantsev2010}, the three integrals were calculated numerically. In the case of circular in-plane symmetry, $\alpha_x=\alpha_y=\alpha$, they can be calculated analytically. Defining $\kappa=\frac{\alpha}{\alpha_z}$ ($\kappa$ measures the aspect ratio: It is small for a flat dot, large for an elongated dot), we obtain
\begin{eqnarray}\label{IxIxIz}
I_x=I_y&&=\sqrt{2} \pi \alpha^\frac{3}{2}\frac{1}{2}\frac{1}{\kappa-1}\left[\sqrt{\kappa}-\frac{\sin^{-1} \sqrt{1-\kappa}}{\sqrt{1-\kappa}}\right]\nonumber\\
I_x=I_y&&=\sqrt{2} \pi \alpha^\frac{3}{2}\frac{1}{2}\frac{1}{\kappa-1}\left[\sqrt{\kappa}-\frac{\sinh^{-1} \sqrt{\kappa-1}}{\sqrt{\kappa-1}}\right],\nonumber\\
\end{eqnarray}
and
\begin{eqnarray}\label{IxIxIz}
I_z&&=\sqrt{2} \pi \alpha^\frac{3}{2}\frac{1}{\kappa-1}\left[\frac{\sin^{-1} \sqrt{1-\kappa}}{\sqrt{1-\kappa}}-\frac{1}{\sqrt{\kappa}}\right]\nonumber\\
I_z&&=\sqrt{2} \pi \alpha^\frac{3}{2}\frac{1}{\kappa-1}\left[\frac{\sinh^{-1} \sqrt{\kappa-1}}{\sqrt{\kappa-1}}-\frac{1}{\sqrt{\kappa}}\right],\nonumber\\
\end{eqnarray}
for $\kappa<1$ and $\kappa>1$, respectively, in both cases.

\begin{figure} []
\centering
\includegraphics [width=\columnwidth]{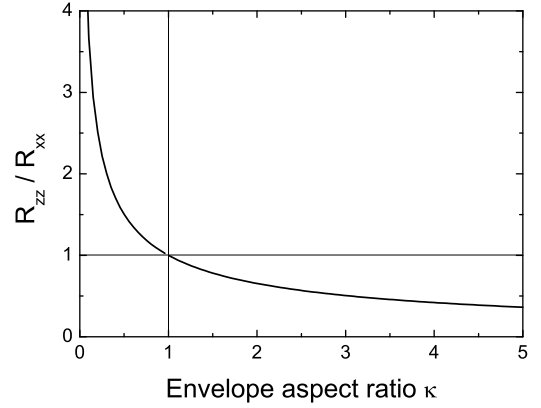}
\caption{Ratio $R_{zz} / R_{xx}$ of the parameters of long-range exchange, as a function of the aspect ratio of Gaussian envelope functions.} \label{fig11}
\end{figure}

The plot of $\frac{R_{zz}}{R_{xx}}$ (=$\frac{I_z}{I_x}$), Fig.~\ref{fig11}, shows a steady decrease when increasing the value of $\kappa$. Note that the ratio reaches unity at $\kappa=1$, which marks an isotropic envelope function for the LH (a point that is however not reached exactly for the isotropic QD). In a flat dot, $R_{zz}$ may be much larger than $R_{xx}$ and $R_{yy}$, implying that, according to Eq.~\ref{delta2}, we expect $\delta'_0$ to be larger than $\delta_0$. Note that the parameters $R_{xx}$ and $R_{yy}$ which govern the position of the $\sigma$-emitting excitons are expected to differ slightly for the LH and HH since they are evaluated with the corresponding envelope functions.

To sum up about the shift of the bright states with respect to the dark state of the LH exciton, the ratio of the $\pi$-emitting to $\sigma$-emitting shifts is 4 for short-range exchange, but long-range exchange makes it larger in a flat dot (under tensile strain) and smaller in an elongated dot (under compressive strain).

Finally, it is useful to write the restriction of the Hamiltonian to the LH excitons in the basis used in the main text, \emph{i.e.}, the conduction electron / valence electron states based on Eq.~\ref{States}:

\begin{equation} \label{ehLH}
\begin{pmatrix}
|+0\rangle& |+1\rangle& |-1\rangle & |-0\rangle\\
\hline
\frac{2}{3}  \omega_\pi  & 0 & 0 & \frac{2}{3}  \omega_\pi  \\
0 & \frac{1}{3}  \omega_\sigma  & 0 & 0 \\
0 & 0 & \frac{1}{3}  \omega_\sigma  & 0 \\
\frac{2}{3}  \omega_\pi  & 0 & 0 & \frac{2}{3}  \omega_\pi
\end{pmatrix}
\end{equation}

Two different parameters $\omega_\sigma$ and $\omega_\pi$ have been introduced following the above discussion (in the model, $\omega_\sigma=\delta_0$ and $\omega_\pi=\delta'_0$). The eigenenergies are now 0 (dark state), $\frac{\omega_\sigma}{3}$ and  $\frac{4\omega_\pi}{3}$.

\section{Material parameters} \label{app:MatPara}

\begin{itemize}
  \item The numerical calculations have been performed for a CdTe QD in a ZnTe shell. The lattice parameters, elastic, and dielectric constants of the materials are: $a_0 = $6.481 {\AA},
$c_{11} = $61.5 GPa, $c_{12} = $43 GPa, $c_{44} = $19.6 GPa,
$\varepsilon_r = $10.6 for CdTe, and $a_0
= $6.104 {\AA}, $c_{11} = $71.6 GPa, $c_{12} = $40.7 GPa, $c_{44} =
$31.2 GPa, $\varepsilon_r = $10.1 for
ZnTe. The Luttinger parameters, spin-orbit energy and deformation
potentials (Bir and Pikus Hamiltonian \cite{BirPikus}) are: $\gamma_1 = $4.6, $\gamma_2 = $1.6, $\gamma_3 =
$1.8, $\Delta = $0.9 eV, $a_v = $0.55 eV, $b = $-1.23 eV, $d = $-5.1
eV for CdTe and $\gamma_1 = $4.07, $\gamma_2 = $0.78, $\gamma_3 =
$1.59, $\Delta = $ 0.95 eV, $a_v = $0.79 eV, $b = -$1.3 eV, $d =
-$4.3 eV for ZnTe.
  \item GaAs \cite{Vurgaftman2001}: $c_{11} = $120 GPa, $c_{12} = $55 GPa, $b$=-2.0eV, $\Delta_{SO}$=0.35eV, $a_0$=0.565nm
  \item InAs: $c_{11} = $92.2 GPa, $c_{12} = $46.5 GPa, $c_{44} = $44.4 GPa, $d$=-3.6eV, $\Delta_{SO}$=0.39eV, $a_0$=0.606nm
  \item InP: $a_0$=0.587nm
  \item Ge: $c_{11} = $126 GPa, $c_{12} = $44 GPa, $c_{44} = $67.7 GPa, $d$=-5.28eV, $\Delta_{SO}$=0.29eV, $a_0$=0.566nm
  \item Si: $c_{11} = $166 GPa, $c_{12} = $64 GPa, $b$=-2.1eV, $\Delta_{SO}$=0.044eV, $a_0$=0.543nm
\end{itemize}

\vspace{0cm}

\end{document}